\documentclass[10pt,journal,compsoc]{IEEEtran}
\ifCLASSINFOpdf
  \usepackage[pdftex]{graphicx}
  \graphicspath{{Figures/}}
\else
  \usepackage[dvips]{graphicx}
  \graphicspath{{Figures/}}
\fi
\ifCLASSOPTIONcompsoc
  \usepackage[nocompress]{cite}
\else
  \usepackage{cite}
\fi
\usepackage{amsmath}
\usepackage{csvsimple}
\usepackage{color}
\usepackage[hyphens]{url}
\usepackage{nicefrac}
\usepackage{siunitx}
\usepackage{array,framed}
\usepackage{booktabs}
\usepackage{caption, subcaption}
\usepackage{verbatim}
\usepackage[normalem]{ulem}

\newcommand{\mg}[1]{{\color{black}{#1}}}

\newcommand{\new}[1]{{\color{black}{#1}}}

\def\predicio{\texttt{GT-DB} }
\def\netacuity{\texttt{GeoIP-DB-A} }
\def\maxmind{\texttt{GeoIP-DB-B} }

\begin{document}
\title{A deep dive into the accuracy of IP Geolocation Databases and its impact on online advertising}

\author{Patricia Callejo,
        Marco~Gramaglia,
        Rub\'en~Cuevas, and
        \'Angel~Cuevas
\IEEEcompsocitemizethanks{\IEEEcompsocthanksitem Authors are with University Carlos III of Madrid.

\IEEEcompsocthanksitem P. Callejo, R. Cuevas, and A. Cuevas are also with UC3M-Santander \\Big Data Institute.
\IEEEcompsocthanksitem Corresponding e-mail: patricia.callejo@uc3m.es
}
} 

\markboth{IEEE Transactions on Mobile Computing}%
{P. Callejo \MakeLowercase{\textit{et al.}}: A deep dive into the accuracy of IP Geolocation Databases and its impact on online advertising}
\IEEEtitleabstractindextext{%

\begin{abstract}
The quest for every time more personalized Internet experience relies on the enriched contextual information about each user.
Online advertising also follows this approach. Among the context information that advertising stakeholders leverage, location information is certainly one of them. However, when this information is not directly available from the end users, advertising stakeholders infer it using geolocation databases, matching IP addresses to a position on earth. The accuracy of this approach has often been questioned in the past: however, the reality check on an advertising stakeholder shows that this technique accounts for a large fraction of the served advertisements.
In this paper, we revisit the work in the field, that is mostly from almost one decade ago, through the lenses of big data. More specifically, we, $i$) benchmark two commercial Internet geolocation databases, evaluate the quality of their information using a ground-truth database of user positions containing over 2 billion samples, $ii$) analyze the internals of these databases, devising a theoretical upper bound for the quality of the Internet geolocation approach, and $iii$) we run an empirical study that unveils the monetary impact of this technology by considering the costs associated with a real-world ad impressions dataset.

\end{abstract}
}

\maketitle

\IEEEdisplaynontitleabstractindextext

\IEEEpeerreviewmaketitle

\ifCLASSOPTIONcompsoc
\IEEEraisesectionheading{\section{Introduction}\label{sec:intro}}
\else
\section{Introduction}
\label{sec:intro}
\fi
\IEEEPARstart{T}{he} ubiquitous connectivity provided by modern cellular technologies, and the success of the mobile web and application paradigms, introduced the necessity to contextualize the service provided with location information. While smartphones have supported this capability since their infancy, the complexity of the World Wide Web (which is the common back end for practically the entire landscape of mobile applications) and the growing concerns on the user privacy requirements, makes the structured gathering of such information difficult. For instance, geographic extensions of HTTP headers were proposed~\cite{http-geo-header} but never approved in IETF, leaving this information only at application level either through JavaScript~\cite{js-geo}
or with OS APIs, upon user permission. \new{Hence, user devices such as mobile phones only share precise positioning data (such as the one got by GPS devices) if they are explicitly configured to do so, and only with applications and servers that have got a specific user permission.}

\new{Still, locating users and terminals can also be useful outside the application domain, so network operators and third-party developers are constantly using alternative technologies to achieve location knowledge. For instance, network operators (that have access to information coming from the lower layers of the network) can reconstruct users positions and trajectories by inferring them through the visited cell-towers~\cite{fiore2020privacy}. However, the most used technique to get positioning information without the explicit gathering of GPS data is IP Geolocation, GeoIP for short. This practice is very common, and it is used for important tasks in the online services landscape, including geofencing~\cite{rodriguez-garzon-geofencing}, fraud detection, and online advertising~\cite{iab_location_1,iab_location_2}, our focus.}

As demonstrated by recent studies~\cite{10.1145/2987443.2987451,10.1145/2740908.2742728,VANDAM201560,ESTRADAJIMENEZ201732,10.1145/2716281.2836098}, data brokers and ad-tech providers track and create profiles from user activities using any kind of data items 
and the physical location of end users is just one of them. 
It is important to note that GeoIP is by far the most employed methodology. According to our data, coming from an online advertising stakeholder, \new{which we will describe in detail in \S \ref{sec:dataset}}, at least 50\% of the handled ad-requests by an online advertising stakeholder included a user location inferred through an IP Geolocation Database, \new{GeoIP Database for short}.

Physically pinpointing Internet hosts on the earth is a problem that has been studied in the early 2000s~\cite{padmanabhan-geomapping} with active measurement technologies, and several \new{GeoIP} databases have been available for getting the latitude and longitude information given an IP address for over ten years. However, their precision has often been disputed~\cite{poese2011ip} since basically their beginnings. In contrast to the literature, which we thoroughly review in \S \ref{sec:relwork}, in this work, we tackle this problem from a different perspective. \new{For first time we analyze the performance of GeoIP databases within the context of a business use case, i.e.,  online advertising. The larger scale and richness of our dataset allows to contribute the first upper-bound of GeoIP performance as well as adding} more insights to the state-of-the-art work already available on this subject:

\noindent \textbf{- Ground-truth:} the extensive data we analyze contains the ground-truth position for each of the mobile terminals (e.g., smartphones or tablets) as well as their associated IP addresses. This precise ground-truth location, gathered by using high precision Geolocation technologies such as GPS, allows us to compute the error that \new{GeoIP} generates compared to the actual ground-truth position of the devices, without resorting to other kind of approximations for the real location of the IP address of the end users. \\

\noindent \textbf{- Pervasiveness:} previous studies usually resort to either a few number of devices that can simultaneously record the IP address and the position of the terminal, or limit their study to a subset of hosts, whose position is known \textit{a priori}. Instead, the data we use in this study covers millions of mobile terminals over \new{three} countries \new{(Spain, France and Great Britain)}  for one month. To the best of our knowledge, this is the first study that unveils the performance of GeoIP at this scale. 

\new{\noindent \textbf{- Heterogeneity:} previous works usually focus on a specific network access technology: either fixed hosts using cabled access, or mobile hosts connecting to the internet through wireless mobile broadband. The dataset we analyze in this study contains both types of connectivity, gathered through the analysis of mobile phone data. Namely, we study WiFi connectivity (hence covering IP addresses assigned by network operator to fixed internet access) and cellular ones (fully covering IP addresses assigned by mobile network operators). This allows us to assess the performance of GeoIP for many use cases, and detailing the monetary implications of this technology for online advertising. }

\noindent \textbf{- Deepness:} besides evaluating the performance of GeoIP, thanks to the extensiveness of the datasets we analyze in this work, we are able to provide deeper insights and draw possible theoretical upper bounds for the performance of \new{GeoIP}, which could be substantially improved, according to our analysis.

\noindent \textbf{- Impact on technology:} motivated by the extreme usage of GeoIP in the context of online advertising, we run an empirical evaluation to assess the convenience of using GeoIP databases in ad campaigns from a budgetary point of view. Contrary to the conventional wisdom, our results show that despite the obvious lack of accuracy of current \new{GeoIP} Databases, the potential higher cost of more precise technologies, such as GPS, may make GeoIP the most economically efficient location technology to be used under certain configurations of ad campaigns.
    
\new{As discussed above, in this paper we analyze the effectiveness of two major GeoIP databases, providing these specific contributions:

\begin{itemize}

    \item We revisit the findings of works published around one decade ago on the precision of GeoIP leveraging on a large scale ground-truth dataset. We assess the performance of GeoIP for several metrics, and quantify their best theoretical performance.

    \item  We analyze the effectiveness of GeoIP databases when dealing with different use cases. In particular, we analyze the monetary implications of GeoIP in the context of online advertising.

\end{itemize}

}

\new{The paper is structured as follows:} in \S \ref{sec:background} we discuss the main motivation of this work: location-based advertising and its ecosystem. In \S \ref{sec:dataset} we introduce the large scale datasets we used to evaluate the performance of two largely used GeoIP databases, first by assessing their quality under different scenarios (\S \ref{sec:accuracy}), then by analyzing their internals (\S \ref{sec:voronoi}), and eventually by quantifying their impact on real-world online advertising campaigns (\S \ref{sec:advertising}). We finally position our work in the state-of-the-art in \S \ref{sec:relwork} before concluding in \S \ref{sec:conclusion}.

\section{Background}
\label{sec:background}

In this paper, we study the performance of GeoIP databases in their usage in online advertising. To this end, in this section, we briefly summarize the background on these topics. 

\subsection{\new{GeoIP} Databases}
\label{sub:geoipdb}

\new{A GeoIP Database} provides the mapping between any IP address in the world to a \texttt{lat,long} coordinate. 

\new{In contrast with other options, such as gathering user GPS data, GeoIP has a very high scalability and pervasiveness, as $i$) an IP address labels every host in the Internet (even the ones that do not have a GPS device active), and $ii$) it is an information available to any network element, with no specific permission granted by the end user.}
Hence, GeoIP databases are arguably the only geolocation technology that meets the needs of scale and coverage \new{for businesses such as online advertising, fraud detection, or antipiracy.}

For these reasons, despite their accuracy has always been questioned, it is also a reality that GeoIP databases are the \emph{de facto} most used technology for location-based services on the Internet.

\new{The details on the specific IP to location algorithms used by free and commercial~\cite{maxmind,netacuity, ip2location, neustar} databases are often not disclosed, and range from very simple WHOIS lookups up to measurements on the delay associated to an address from different vantage points on the network infrastructure.}
However, the technical information shared by some providers \cite{maxmind,netacuity} as well as previous academic papers studying them \cite{poese2011ip,gouelip}, 
agree on the fact that GeoIP databases \new{are built following a common approach. Providers} divide the space of IP addresses into autonomous systems and further split them into variable sized IP prefixes. Then, by using active and passive measurements, they map the overall set of prefixes onto a geographical grid of \emph{anchor points}, that is their best estimation of the position of each prefix. We will also use this assumption throughout the paper.

\subsection{Location Data in online advertising}
\label{subsec:loc_data_back}

The differentiation factor of online advertising compared to other forms of advertising is its capacity to perform fine-grained ad targeting campaigns based on audiences defined by the targeted users' demography (e.g., age and gender), preferences and interests (e.g., sports, restaurants, etc.), and \emph{location} information. 
Location is then a fundamental parameter in the definition of online advertising campaigns~\cite{iab_location_1,iab_location_2}.

\subsubsection{\new{A primer on online advertising}}
\label{subsec:primer}
Advertisers configure their campaigns in technological platforms referred to as Demand Side Platforms (DSPs), which receive offers of available ad spaces from tens of thousands of different publishers through Ad Exchanges (AdX).

The AdX maps each ad-request from a publisher into a bid-request message which is sent to several DSPs. Each DSP checks if the properties of the ad-request (e.g., user's demographic, interests and location) match the configuration parameters of any of its campaigns. If so, the DSP returns a bid-response, including the price it is willing to pay for the offered ad space. The AdX runs a real-time auction process based on the received bid-responses and selects the winning DSP, which will handle the delivery of the ad impression to the user. 

The ad delivery process is (obviously) subject to a monetary transaction. The most common pricing schemes in online advertising are CPM (Cost per one thousand impressions) and CPC (Cost Per each click on an ad impression). Note that CPM and CPC are metrics that are known a posteriori, once the campaign is finished. A proxy metric for the cost of an ad impression is the \emph{bid floor}. This is a variable in the bid-requests that indicates the minimum bidding price accepted by the publisher offering the ad space. 

\subsubsection{\new{Location data sources}}
\label{subsec:locdatasources}

DSPs have access to the location associated to an ad space through the location information embedded in bid-requests \new{from three possible data sources~\cite{openrtb}.}

\noindent \textbf{- User}: The location data is provided by the user and embedded in the ad-request. For instance, location information (e.g., an address) provided by the user through a registration form. This type of location appears rarely in bid-requests.

\noindent \textbf{- GPS/Location Services}: This type of data is expected to provide high-precision, and in practice, it should directly come from the positioning device of the user, offering GPS precision. Given the high-precision of the data, bid-requests, including this type of location data, are expected to have a higher starting bid price for the auction.

\noindent \textbf{- IP address}: An important number of ad-requests leave the user device without any location information. Due to the importance of location in online advertising, it is common that one of the intermediaries in the ecosystem (e.g., the AdX) enriches the ad-request or its correspondent bid-request with location information based on the IP address of the device. To this end, they use GeoIP databases described above. 

To understand the importance of GeoIP in the online advertising ecosystem, we have computed the fraction of daily bid-requests including a GeoIP, \new{GPS or unavailable} location received by TAPTAP Digital~\cite{taptap}, a mid-size DSP \new{(See details in \S \ref{sec:dataset})}, in its bid stream (i.e., the bid-requests flow). \new{In particular, we have measured this metric for the bid-requests of the three countries analyzed in this paper (Spain, France, and Great Britain) during a period of 16 days. The results show that the average fraction of daily bid-requests across the considered countries, including GeoIP, GPS or unavailable location data is 52\%, 18\% and 30\%, respectively. In particular,  48,0\%, 50,8\%, and 57,3\% of the bid requests for Spain, France, and Great Britain, include GeoIP location information, respectively. It is important to remark that most DSPs process bid requests with unavailable location data. They extract the IP address of the device from the bid request and obtain an associated location from a GeoIP database. This location data assignation technique allowS DSPs to effectively have a location for all received bid requests. In summary, roughly half of the bid-requests (and up to 80\% in those DSPs using the described location data assignation technique) include locations extracted from GeoIP based on our dataset.} These values further corroborate the fact that GeoIP is \new{the most common} technology for providing location information in online advertising. 

If an ad-request does not include a location context, depending on the kind of ad campaign, it will be unlikely to find a matching user. Hence, location is definitely a very sensitive parameter for the efficiency of an ad campaign, which may range from coarser levels (i.e., country) to very fine ones, targeting users at a zip code level.

\subsection{Ground-truth location data}
\label{subsec:loc_services}

To achieve our goal of assessing the accuracy of GeoIP location data and its impact on online advertising, we had to resort to a data source that provides high-precision geolocation information for an extremely high volume of users. 

Multiple location providers collect high-precision location information from users. Some examples are Safegraph~\cite{safegraph}, Cuebiq~\cite{cuebiq}, Foursquare~\cite{foursquare}, and Tamoco~\cite{tamoco} to name a few. These providers use different techniques to obtain accurate location information from users, as described next:

\noindent \textbf{- Embedded SDKs in mobile apps}: The location provider agrees to include its SDK in the mobile app(s) of a given app developer. This SDK leverages the permission granted by the end users to the ``host'' application and collects the GPS location information from the device, as well as other parameters, including the IP address.

\noindent \textbf{- Check-ins}: The user proactively registers a check-in at a specific venue (e.g., restaurants, coffee shops, etc.) when it happens (usually to contextualize posts on online platforms), the accurate location of the venue is well-known, and thus users can be located with high-precision and with total transparency to them, as they are consciously interacting with the app to provide such information. 

In this paper, we use a dataset \new{from a location provider distributing an embedded SDK in mobile apps (see details in \S \ref{sec:dataset})} as our ground-truth information for the precise positioning of end users.

\section{Datasets}
\label{sec:dataset}

This section describes the datasets and the evaluation scenarios we use in the remainder of the paper. In our study, we limit the analysis to three major European countries (Spain, France, and Great Britain) where online advertising presents a strong penetration and for which we have a good coverage in our datasets.

\subsection{\new{GeoIP} Databases}
\label{sub:geoipdata}
We leverage two of the most widely used \new{GeoIP} databases to analyze the performance of the GeoIP location technology. We keep the name of these providers anonymous since our research aims not to scrutinize specific providers but rather assess the performance of the GeoIP in the context of online advertising. In the rest of the paper, we name these datasets that refer to these GeoIP databases as \netacuity and \maxmind, respectively. Both providers offer their database as commercial products, have wide coverage in the three considered countries, and update their database weekly.
We collected regular snapshots of such databases to check the consistency of the data along the time.
For instance, \netacuity includes all the IP-location samples between July-2020 and May-2021. For \maxmind, instead, we gathered data for the period April-2021 and May-2021. These databases include all the information needed to match the IP address to a position and other side information such as the kind of access technology associated with a given IP.

\subsection{Bid stream dataset}
\label{sub:bidsdata}

In this work, we measure the impact of the accuracy of GeoIP for location-based online advertising by analyzing real bid-request flows (a.k.a. bid stream) gathered from the Sonata DSP~\cite{sonata} operated by TAPTAP Digital~\cite{taptap}, a digital marketing company operating in 15 countries. Sonata is a mid-size DSP whose bid stream includes a large-scale sample of bid-requests generated from Spain, France, and Great Britain. In particular, we processed the bid stream collected by Sonata between 1-May-2021 and 17-May-2021, which includes an average number of daily bid-requests of 257.6M, 64.5M, and 54.1M for ES, FR, and GB, respectively.
While a bid-request may include several user context related features, we only process the fields that are relevant for our study, namely: \texttt{$<$timestamp; IP address; Location Source; lat,long$>$}. The location source field corresponds to those defined in 
\new{\S \ref{subsec:locdatasources}:} GPS, GeoIP, or User, or \emph{unavailable} in case no source is reported. \new{Note that for the analysis we only select the GPS information, as we can use it as ground-truth information.} 

\subsection{(Ground-truth) GPS location data}
\label{sub:gtdata}
To validate the performance of GeoIP, we use a dataset from \new{a location provider distributing an embedded SDK in mobile apps.\footnote{The name of this provider is kept anonymous due to its express request.} This dataset reports GPS location coordinates, which we consider as the reference ground-truth position for the end users. This location data provider operates in more than 15 international markets including Spain, France, Italy, Great Britain, US, Mexico, Argentina, Colombia, South Africa, etc. Its SDK is embedded in dozens of applications, including popular applications such as weather, news or radio apps. It offers a coverage of at least 5\% of the population in the main markets where it operates. Finally, our provider's  location data is used by customers across different sectors such as  online advertising, retail, e-commerce, real state and financial services, among others.}

In particular, this dataset includes the following data tuple per location event: \texttt{$<$timestamp; lat,long; IP address; carrier$>$}. The dataset spans a period of 30 days (from 1-Sep-20 to 30-Sep-20) and provides a very reliable snapshot of the mobile users. On average, the number of daily location samples is 31M, 16M, and 20M for ES, FR, and GB, respectively. In total, we have for the three countries more than 2.05B data samples for the considered period. To the best of the authors knowledge, this is the largest ground-truth dataset ever used for analyzing the performance of \new{GeoIP}  databases, increasing in several orders of magnitude the datasets used in previous studies. We refer to it as \predicio in the remainder of the paper. 
\new{As we report in \S \ref{subsub:maxmind_validation}, the results obtained with the ground-truth data from our provider are aligned with those reported by a major GeoIP Database provider. We believe that this represents a significant hint about the quality of the used ground-truth data.}

\section{GeoIP performance}
\label{sec:accuracy}

This section evaluates the performance obtained by \new{GeoIP} under several scenarios relevant to the online advertising market. We present the overall methodology implemented to compute GeoIP performance in \S \ref{sub:accmetho}, before analyzing the different results in the following subsections.

\begin{figure*}[ht!]
     \centering
     \begin{subfigure}[b]{0.28\textwidth}
         \centering
         \includegraphics[trim=0cm 2.1cm 0cm 2cm, clip,height=\textwidth]{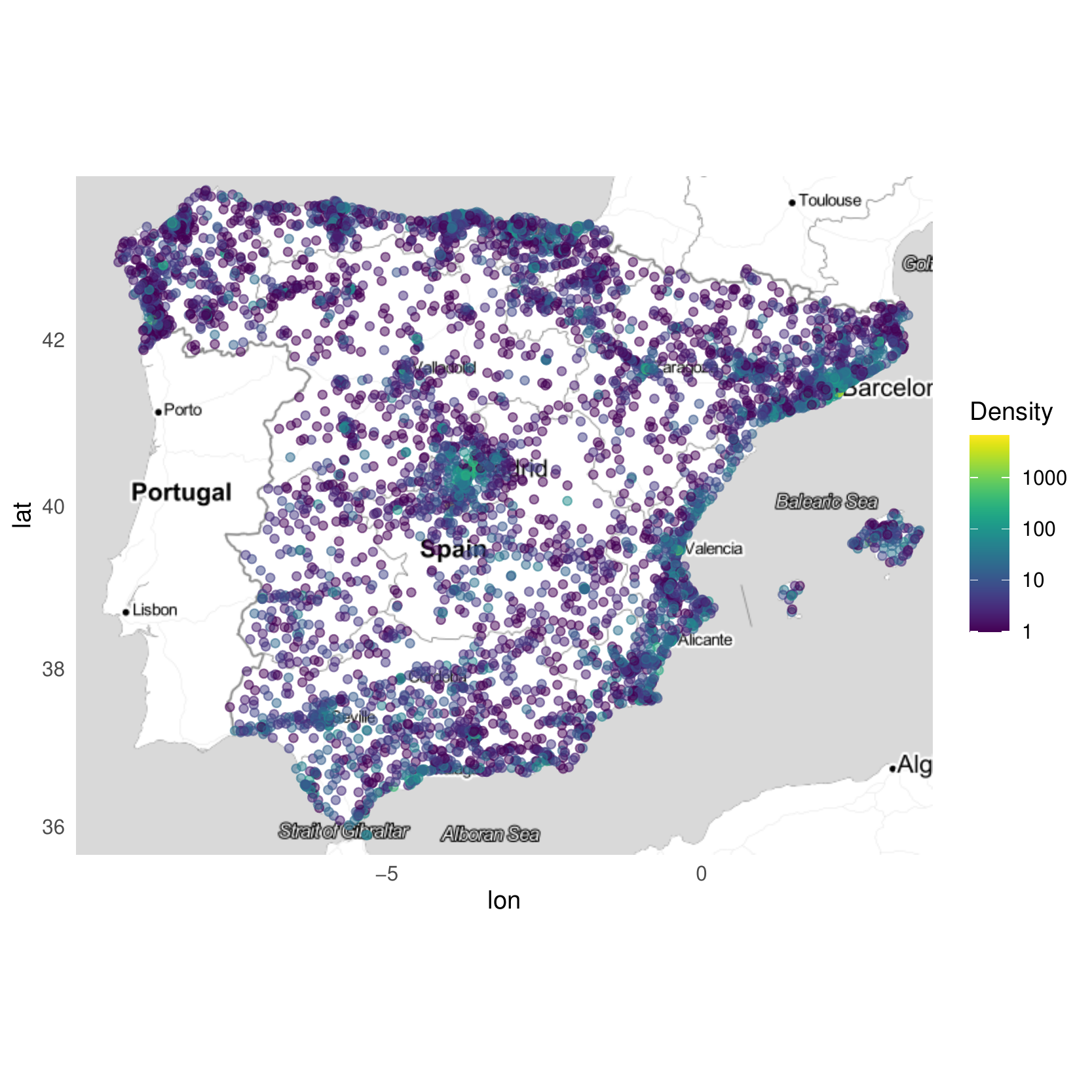}
         \caption{Spain}
         \label{fig:map_spain}
     \end{subfigure}
     \hfill
     \begin{subfigure}[b]{0.28\textwidth}
         \centering
         \includegraphics[trim=0cm 1.8cm 0cm 1.8cm, clip,height=\textwidth]{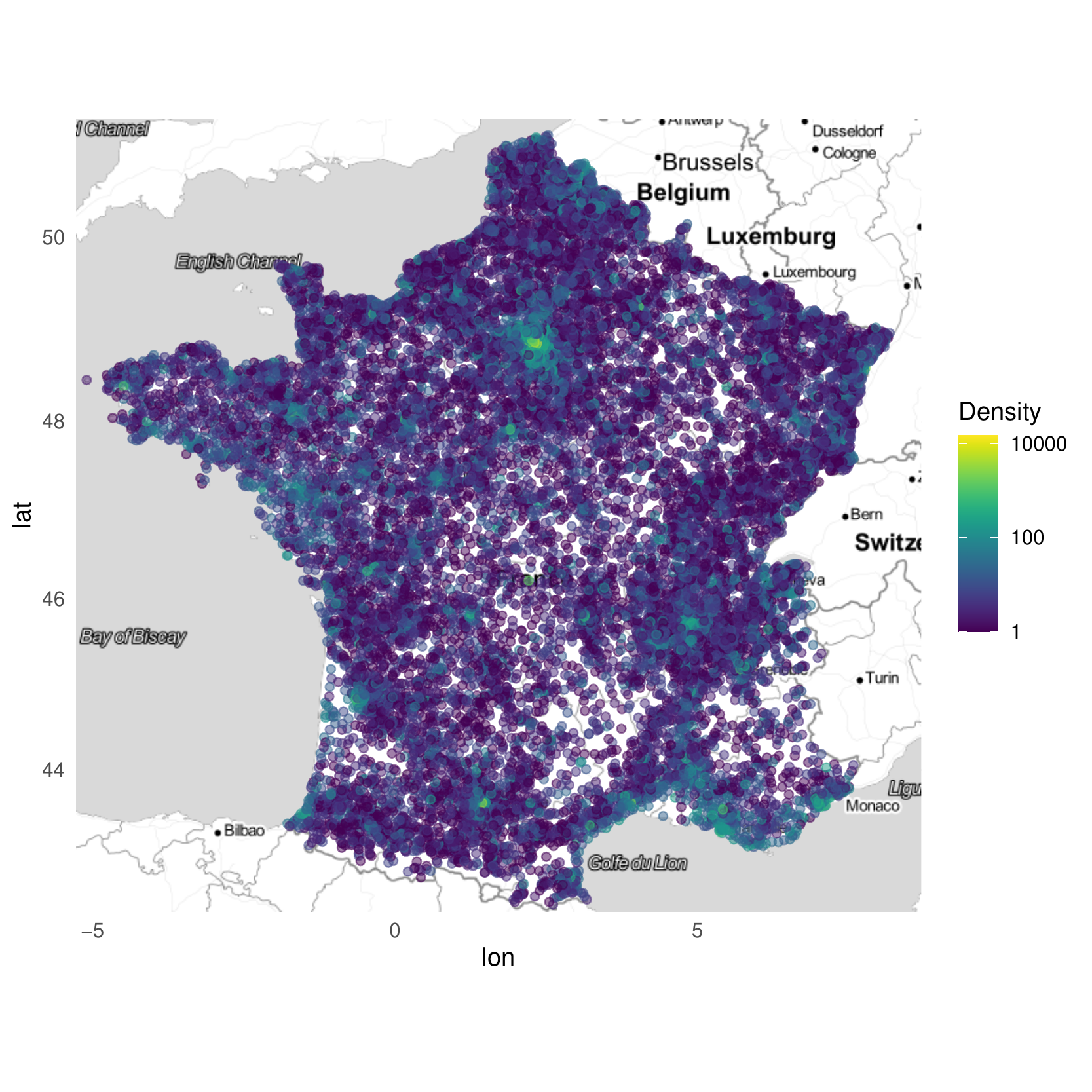}
         \caption{France}
         \label{fig:map_fr}
     \end{subfigure}
     \hfill
     \begin{subfigure}[b]{0.28\textwidth}
         \centering
         \includegraphics[trim=1.5cm 0cm 1.6cm 0cm, clip,height=\textwidth]{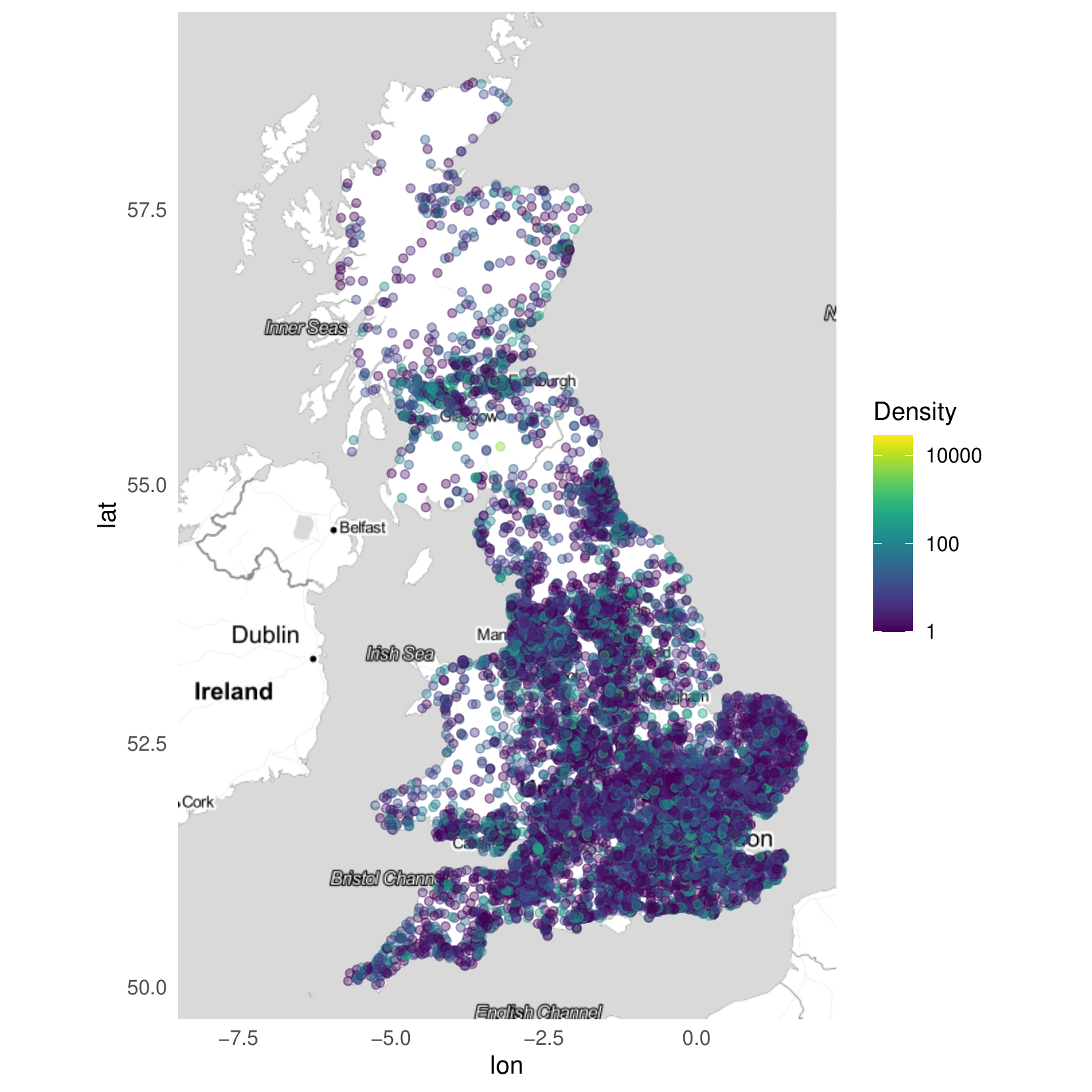}
         \caption{Great Britain}
         \label{fig:map_uk}
     \end{subfigure}
        \caption{\new{Number of anchor points in the analyzed countries. Brighter colors indicate a larger concentration of anchors. (Figure best viewed in colors).}}
        \label{fig:map}
\end{figure*}

\subsection{Methodology}
\label{sub:accmetho}

\begin{table}[t]
\resizebox{\columnwidth}{!}{%
\begin{tabular}{@{}llll@{}}
\toprule
                 & \textbf{Spain}       & \textbf{France} & \textbf{Great Britain}              \\ \midrule
\textbf{Level 1} & Post Code            & Post Code       & Post Code                \\ 
\textbf{Level 2} & City/Municipality    & Commune         & Local Authority District \\
\textbf{Level 3} & Province             & Department       & County / Region          \\
\textbf{Level 4} & Autonomous Community & Region          & Country                  \\
\textbf{Level 5} & Country              & Country         & Kingdom                  \\\bottomrule
\end{tabular}}
\caption{Administrative Levels considered in each country.}
\label{tab:admin_levels}
\end{table}

We benchmark \netacuity and \maxmind using as reference the \predicio database, which provides high-accuracy location for several millions of users and precise time information that allows us to compare the ground-truth samples to the proper instance of the GeoIP databases.

By joining \predicio with \netacuity and \maxmind we obtain two \emph{latitude, longitude} pairs for the same IP address at a specific time, one belonging to \predicio (used as ground-truth, $pos_{GT}$) and the other belonging to the GeoIP generated instance, $pos_{IP}$. Thus, for all the IP addresses of the \predicio database we compute the distance between $pos_{GT}$ and $pos_{IP}$ using the Haversine distance \cite{haversine} formula, which yields the distance between any \emph{latitude,longitude} pairs on the earth. Formally,
\begin{equation}
\mathcal{E}=\mbox{hav}\left( pos_{GT},pos_{IP}\right)
\end{equation}

We used this approach in the rest of this section for assessing the databases' \textit{Precision}, which is a metric that evaluates the pure distance.

However, ad campaigns usually include specific location targets that often correspond to concrete administrative boundaries, such as countries, regions, cities, or zip codes. Therefore, in the context of online advertising, the performance of a GeoIP service should be measured by its capacity to locate users within the targeted administrative region properly. We refer to this metric as \textit{Accuracy} in this paper.

For the accuracy analysis, we used the \emph{Shapefiles} available on the open data portals~\cite{opendatauk,opendatafr} of the different countries we analyzed and extracted the geographical extent information related to the different administrative regions. In order to increase the scalability of this analysis, we divided the space into a fixed grid using the Uber H3~\cite{h3} geographical spatial index, to transform geographical joins into standard joins.

We formally define the accuracy metric $\mathcal{A}$  as follows:

\begin{equation}
\mathcal{A} = \frac{pos_{IP} \in R \mid pos_{GT} \in R}{pos_{IP} \in R}
\end{equation}

where $R$ is the targeted spatial region associated to e.g., an administrative division. Our accuracy analysis considers 5 different administrative levels from smaller (Level 1) to larger (Level 5) size as reported in Tab.~\ref{tab:admin_levels}\footnote{Note that Levels 1 and 2 are not always hierarchical. For instance, there are some zip codes in rural areas in Spain that include several villages.}.

\subsection{Space and Time variability}
\label{sub:variability}

Before analyzing the performance of GeoIP, we discuss in this subsection \new{some overall statistics of the analyzed GeoIP datasets}.
As introduced in \S \ref{sec:background}, location information is actually inferred based on IP prefixes rather than IP addresses (i.e., contiguous IP addresses usually share the same position). 

\new{We report in Table~\ref{tab:extent} the \predicio extent in the three countries under study, obtained by performing an exhaustive search on the entire IP addresses space. Besides the number of different ranges and anchor points, we also compute the reuse factor, i.e., the number of IP ranges that are mapped to the same position.}

\begin{table}[t!]
\centering
\begin{tabular}{@{}llll@{}}
\toprule
                 & \textbf{Spain}       & \textbf{France} & \textbf{GB}              \\ \midrule
\textbf{\# IP ranges} & 139687              & 399500         & 1051937                  \\
\textbf{\# \mg{anchor points}} & 5288 & 16367          & 10448                  \\ 
\textbf{Reuse factor} & 26.41 & 24.40          & \mg{100.68}                  \\
\bottomrule
\end{tabular}
\caption{Extent of the \netacuity database.}
\label{tab:extent}
\end{table}

The analysis of the reuse factor shows a good correlation with the average population density in the specific countries: Spain and France, with a population density of 92.76  and 123.28 persons per $Km^2$, present a reuse factor around 25 (26.41 and 24.40, respectively). Great Britain, instead, has a much higher population density (279.95 persons per $Km^2$) that is reflected by a higher reuse factor as well, \mg{100.68}.

Fig. \ref{fig:map} depicts the spatial landscape of the $|pos_{IP}|$ set, which reports a similar conclusion. The algorithms implemented by the \netacuity database accurately match the most densely populated areas of Spain, France, and Great Britain, such as the capitals and the most populous cities (e.g., Barcelona, Marseille, and the Liverpool–Manchester Megalopolis). In contrast, they present much lower resolution in rural areas such as Castilla-La Mancha region in Spain, the Massif Central in France, and the Scottish Highlands. 
\new{As we quantify in \S \ref{sub:ruralurban},}
the lack of an \emph{anchor point} for these zones 
\new{introduces} a large error in the location estimation for the (fewer) users located there.

Both \netacuity and \maxmind are periodically updated to account for movement among IP ranges and refine the location estimation according to their algorithm. However, we did not notice any substantial deviation in the computed precision over time. Considering a 30 days time window, the median daily error recorded from the two databases only shown a variance of 1.34 m, 2.51 m, and 11.64 m for Spain, France, and Great Britain, respectively\footnote{\mg{This corroborates the fact that providers are constantly improving their records, as the average deviation of the reported IP prefixes position is much higher, as reported by~\cite{gouelip}}}. For this reason, unless otherwise stated, in the rest of the paper, we analyze a time window of 30 days without distinguishing between weekend, weekdays, day or night, as the time dynamics involved in the update process are probably longer.

\subsection{Global cross-country comparison}
\label{sub:variance}

\subsubsection{Precision}
\label{subsub:prec_countries}
\begin{figure}[t!]
    \centering
    \includegraphics[width=0.9\linewidth]{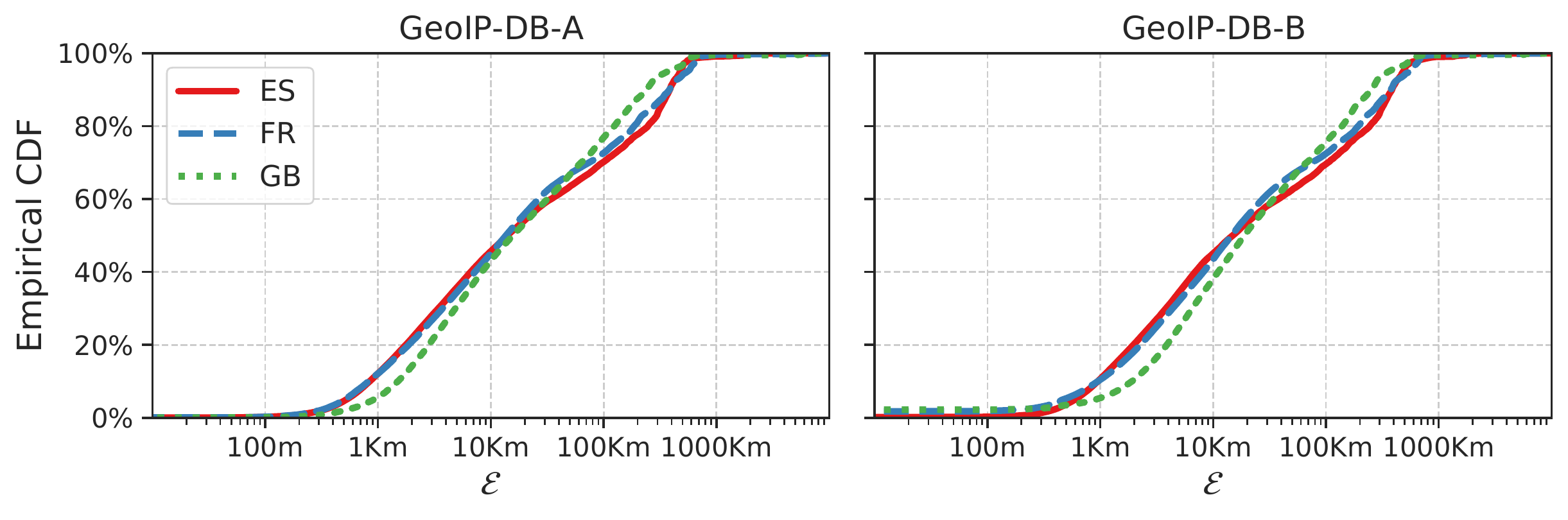}
     
    \caption{Precision across countries}
    \label{fig:cdf_cross_countries}
\end{figure}

We first study the overall precision attained by \netacuity and \maxmind in the reference countries by showing the CDF of $\mathcal{E}$ in Fig.~\ref{fig:cdf_cross_countries}. The precision distribution shows poor behavior in the three countries, and the two explored GeoIP databases, which shows very similar results. For instance, the median error for \netacuity (\maxmind) in Spain, France, and Great Britain is 14.01~Km (14.91~Km), 13.61~Km (14.56~Km), 15.70~Km (18.9~Km), respectively. In addition, there are only a few samples with an error below 1Km (at best 12.1\% for \netacuity in France and 10.6\% for \maxmind in Spain), while the percentage of samples with a very low precision beyond 100~Km grows up to 24\% in the best case (in Great Britain, for \maxmind).

Globally, we notice two main differences in terms of $\mathcal{E}$ among the countries: while Spain and France follow a similar curve, the Great Britain case yields a lower precision when dealing with shorter $\mathcal{E}$ for both databases, compensating with a lower percentage of samples that are associated with very high values of  $\mathcal{E}$. This is quite remarkable when looking at Fig.~\ref{fig:map_uk}, which showcases large areas without any anchor point, showing how the two GeoIP providers can target the potential audience better in GB. 

\begin{figure}[t!]
    \centering
    \includegraphics[width=\linewidth]{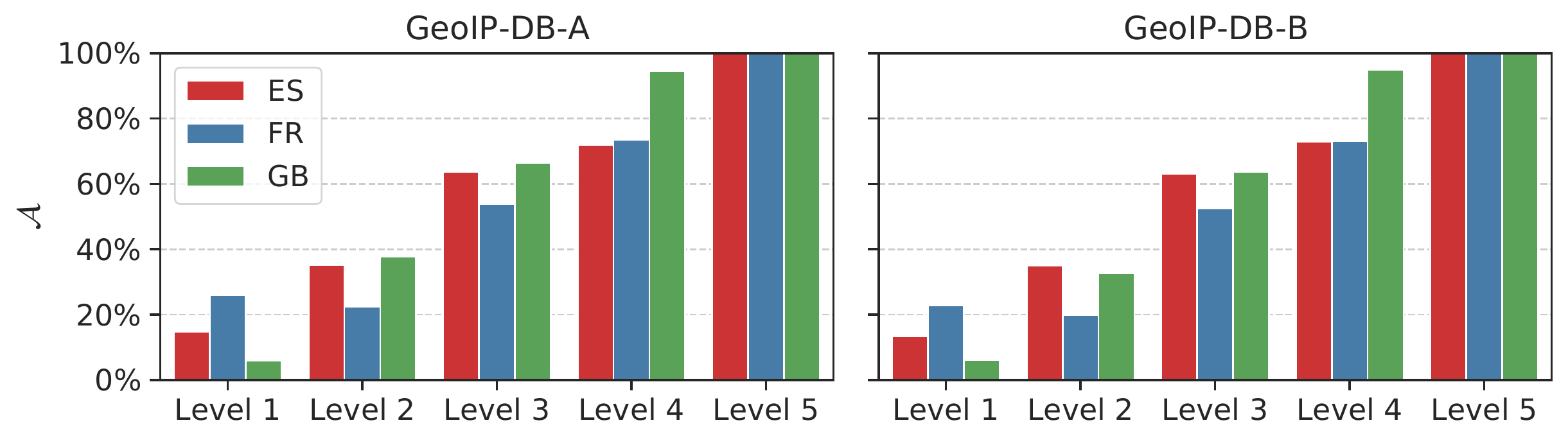}
    \vspace{-5mm}
    \caption{Accuracy by administrative regions}
    \label{fig:accuracy}
\end{figure}

\subsubsection{Accuracy}
\label{subsub:accuracy_countries}

When translating the achieved precision $\mathcal{E}$ into the accuracy $\mathcal{A}$, the possible misplacement has an impact not only depending on \emph{how much} it is, but also on \emph{where} is committed. An error of 5Km between the real and the estimated position may be tolerable if the goal is to locate a user within their province, while it could be too much if the target is a zip code.

Fig.~\ref{fig:accuracy} shows the accuracy results for the two Geolocation IP databases across Spain, France, and Great Britain, and the five administrative regions considered in our analysis (see Tab.~\ref{tab:admin_levels}). First, as expected, the accuracy grows as the size of the considered region increases in all cases. Except for the case of Level 5 (country) where users are correctly located, we observe a quite poor behavior in the remaining levels, which is rather similar in both GeoIP databases. For Levels from 4 to 1, we roughly observe that (at least) 33\%, 40\%, 70\%, and 80\% of the location samples failed to be located in the correct administrative region, respectively. In addition, although the $\mathcal{E}$ distribution is similar across countries, we observe relevant impairments when dealing with accuracy. For instance, while \netacuity can achieve the highest accuracy in France for Level 1 regions (26\%) and the lowest for Great Britain (6\%) their roles are swapped when the task is to locate users within the Level 2 boundaries (22\% for France and 38\% for Great Britain, respectively). 

As the achieved $\mathcal{E}$ and $\mathcal{A}$  by \netacuity and \maxmind are almost equal, in the remainder of the section we focus on the analysis of the performance achieved by \netacuity only. Also, for the accuracy evaluation, we limit the discussion up to Level 4, as Level 5 yields full accuracy.

\subsection{The impact of the urbanization level}
\label{sub:ruralurban}

\subsubsection{Precision}
\label{subsub:prec_ruralurban}

\new{As we anticipated with the discussion of Fig.~\ref{fig:map}, the performance achieved by GeoIP could be quite uneven depending on the location of the real users, because the \emph{anchor points} distribution targets most densely populated areas.
Also, the urbanization level can be a very important factor to consider for different businesses. For instance, should advertisers expect the same level of location precision (accuracy) on the delivered ads from campaigns targeting urban areas vs. rural areas?, how difficult is pinpointing the location of a fraudulent use of a credit card when it is committed from an urban vs. a rural area? Understanding the precision (accuracy) offered by GeoIP across areas with different urbanization levels is key to answer the previous questions.} Thus, following the classification~\cite{classification} provided by the EU countries to distinguish between urban, semi-urban, and rural areas, we categorize $\mathcal{E}$ depending on whether $pos_{GT}$ is in one of the previously mentioned areas.

\begin{figure*}[t!]
    \centering
    \includegraphics[width=0.9\linewidth]{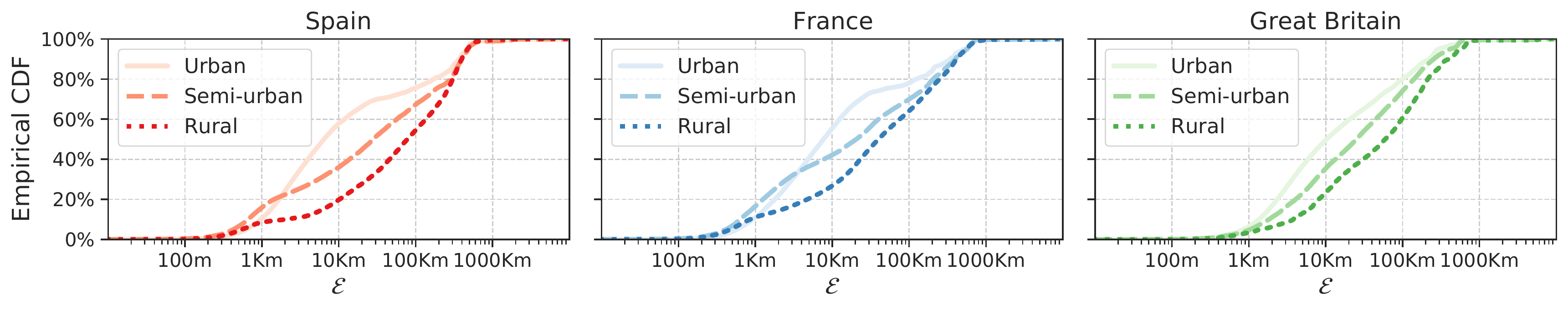}
     \vspace{-4mm}
    \caption{Precision breakdown per urbanization level}
    \label{fig:cdf_rural_vs_urban}
     \vspace{-4mm}
\end{figure*}

The results in Fig.~\ref{fig:cdf_rural_vs_urban} show the precision achieved per country and urbanization level. As in the global country analysis (see \S \ref{subsub:prec_countries}), Spain and France show similar behavior in precision. First, urban areas present a precision $\leq$10Km for 60\% of the samples in both countries. For the same portion of samples, the precision is close to 100Km in both cases for rural areas. It is also interesting to denote that the portion of samples with very high precision ($\mathcal{P} < $ 2 Kms) is higher in semi-urban areas than in urban areas for these countries. Finally, it is worth mentioning that Spain exhibits a higher spread than France between urban and rural areas. In contrast to France and Spain, Great Britain shows a consistent trend in the $\mathcal{E}$ distribution, with a constant 25\% gap between rural and urban areas for all the considered distances.

\new{As we observe in Fig.~\ref{fig:map}, the anchor point distribution is clearly biased towards more densely populated area, a bias that is intrinsically also present in the \predicio dataset. Hence, to further understand the achieved precision depending on the ground-truth location of the users, we compute the correlation between the achieved precision and the number of available anchor points in a given area, using the grid we employ for the accuracy computation.

That is, for every cell in the tessellation we created for the three countries, we first compute the median $\mathcal{P}$ and the count of available anchor points (both on a logarithmic scale), and correlate these two variables with the Pearson's R coefficient. The achieved values (-0.44, 0.53, and -0.57 for Spain, France, and Great Britain respectively) show a very high negative correlation between them: increasing the number of anchor points in a given area (a common circumstance when moving from Rural to Semi-Urban, and from Semi-Urban to Urban) can very likely correspond to a precision improvement of one order of magnitude.

}

\begin{figure*}[t!]
    \centering
    \includegraphics[width=0.9\linewidth]{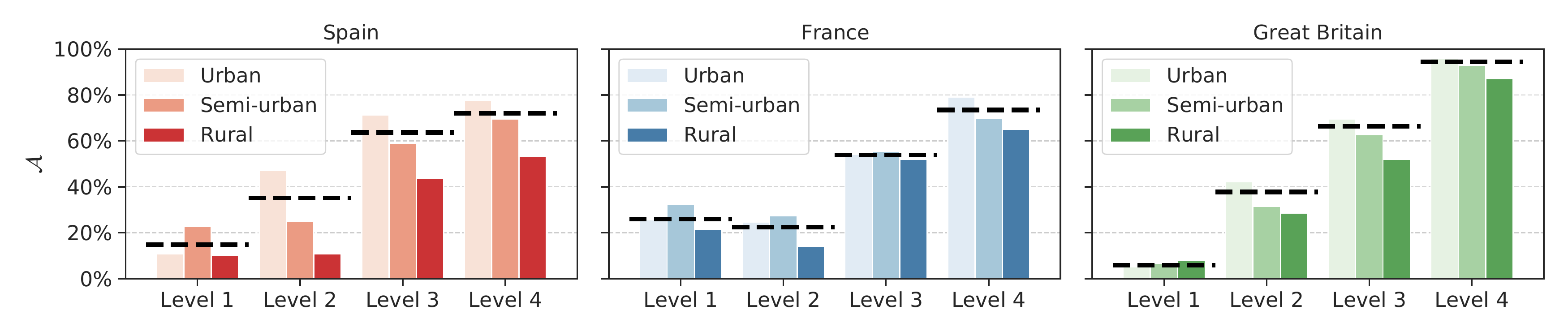}
     \vspace{-4mm}
    \caption{Accuracy by administrative regions and level of urbanization (dashed lines represent the overall accuracy).}
    \label{fig:accuracy_dgurba}
\end{figure*}
\subsubsection{Accuracy}
\label{subsub:acc_ruralurban}

The accuracy $\mathcal{A}$ split across the different administrative levels is presented in Fig.~\ref{fig:accuracy_dgurba}. Spain and Great Britain show significant differences between urban and rural areas in all administrative levels, except Level 1 (i.e., zip code). The Spanish case, in particular, showcases very large differences between $\mathcal{A}$ measured in urban areas and rural areas: for the Level 2 we can observe a dramatic drop from 47\% to 11\%, further corroborating the considerations done in \S \ref{subsub:accuracy_countries}. Contrarily, France presents comparable results among the urbanization degrees, being the least unequal of the analyzed countries. This effect may be because of the more uniform spread of \emph{anchor points} across the country. 

\begin{figure*}[t!]
    \centering
    \includegraphics[width=0.9\linewidth]{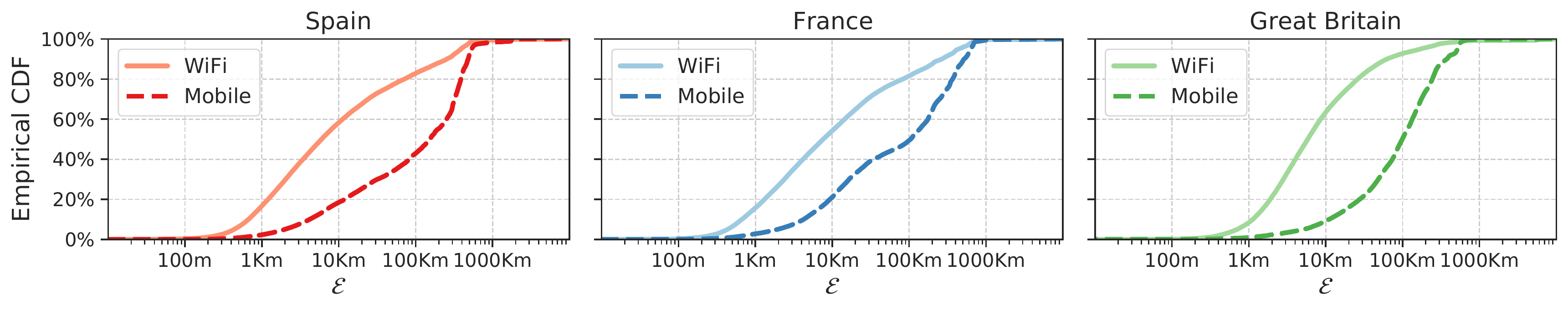}
     \vspace{-2mm}
    \caption{Precision for different access technology: fixed vs. cellular.}
    \label{fig:cdf_connection_type}
     \vspace{-1mm}
\end{figure*}
\begin{figure*}[t!]
    \centering
    \includegraphics[width=0.9\linewidth]{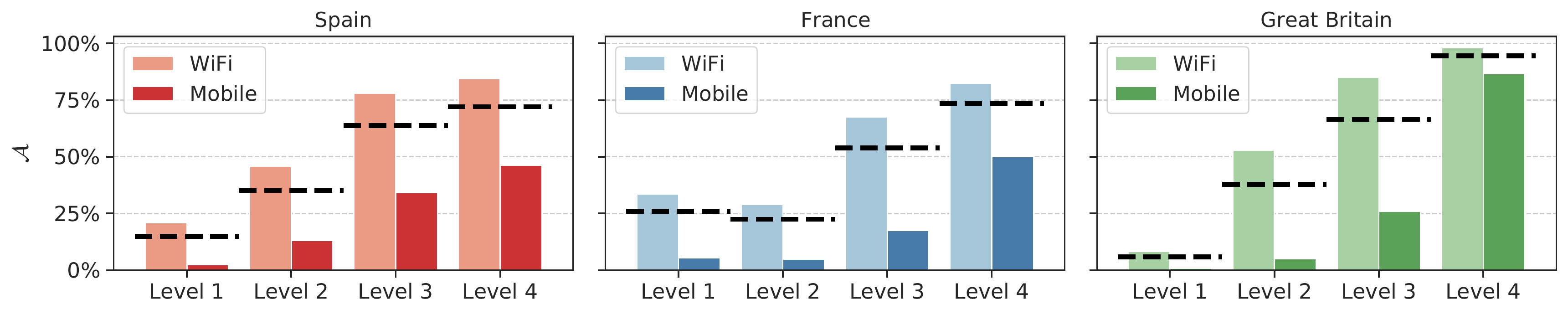}
     \vspace{-2mm}
    \caption{Accuracy by administrative regions and connection type. (dashed lines represent the overall accuracy)}
    \label{fig:accuracy_conntype}
    \vspace{-1mm}
\end{figure*}
\subsection{The influence of the access technology}
\label{sub:access}

\subsubsection{Precision}
\label{subsub:prec_access}
\new{Regardless of the type of devices that are used to gain access to the Internet (e.g., a desktop PC, a laptop, or a mobile phone), a factor that likely affects on the precision of GeoIP services is the access network technology. It seems obvious that pinpointing the location of a mobile device will generate larger errors than estimating the position of a device connected through a broadband fixed-access technology (e.g., Fiber, ADSL, or WiFi).}

Databases such as \netacuity and \maxmind usually offer, for user targeting purposes, also the kind of access technology associated to a given IP address.
However, our ground-truth database \predicio is built using data coming from mobile terminals, which likely only have two types of access network technologies: WiFi and mobile.

\begin{figure*}[t!]
    \centering
    \includegraphics[width=0.95\linewidth]{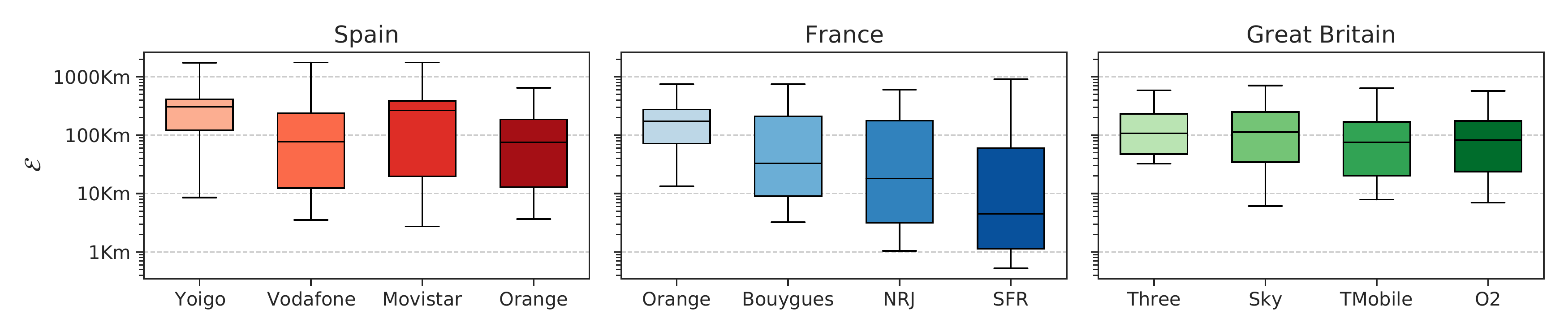}
     \vspace{-2mm}
    \caption{Precision across different ISPs}
    \label{fig:cdf_telcos}
     \vspace{-1mm}
\end{figure*}

\begin{figure*}[t!]
    \centering
    \includegraphics[width=0.9\linewidth]{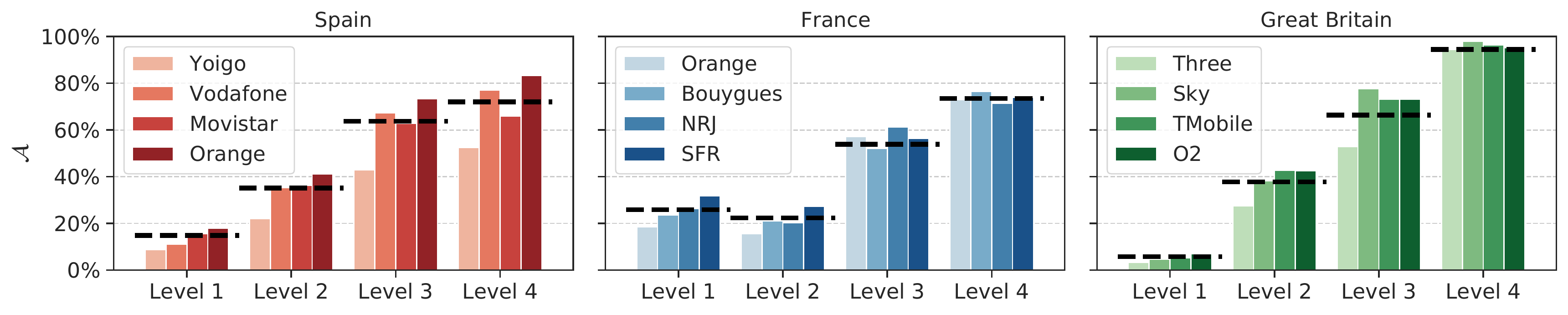}
     \vspace{-2mm}
    \caption{Accuracy by administrative regions and ISPs (dashed lines represent the overall accuracy)}
    \label{fig:accuracy_carrier}
\end{figure*}

Fig.~\ref{fig:cdf_connection_type} shows  $\mathcal{E}$ according to the connection interface inferred from \netacuity for Spain, France, and Great Britain. The behavior is consistent across the countries. As expected, cellular connections lead to much larger errors than WiFi. For most percentiles in the distribution, the gap exceeds one order of magnitude in all the countries. Even more, the number of location samples obtained through the mobile network where $\mathcal{E}\leq$1Km are anecdotal (3\% for the best case, in France). The extremely bad performance of cellular connections is not compensated by WiFi. Even for the best case, in Spain, only 17\% of the users could be located within 1Km, corroborating that this technology, at least for the analyzed databases, is a no-go for precise-location targeted advertising.

\subsubsection{Accuracy}
\label{subsub:acc_access}

It is indeed remarkable that not even the use of the WiFi technology yields good results for the most challenging scenarios: the best case (Level 2 in GB) only achieves $\mathcal{A}$ equal to 52.7\%, while in France this value drops to 29.0\% for the same administrative level. Fig.~\ref{fig:accuracy_conntype} confirms the pronounced unreliability of IP-based geolocation for cellular access technologies, with $\mathcal{A}$ that are often below 10\% for the most challenging scenarios and around 50\% for the least ones (only the Level 4 in GB seems to be well mapped).

\subsection{The variation across different ISPs}
\label{sub:isps}

\subsubsection{Precision}
\label{subsub:prec_isps}

The algorithms employed by \netacuity to map IP addresses to a $pos_{IP}$ may be computed using active latency measures between known milestones on the Internet. Hence, the number and the internal configuration of prefixes for the different ISPs could have a relevant impact on $\mathcal{E}$. We assess this by further split the precision yield by users connected through their mobile interface (discussed in \S \ref{sub:access}) into the different ISPs.
For this purpose, we use the information available in the \predicio database, which collects the carrier name displayed in the mobile terminal. Fig.~\ref{fig:cdf_telcos} shows the achieved $\mathcal{E}$ for the four most relevant carriers in each of the considered countries.

We observe different notable behavior in the yielded $\mathcal{E}$ for all the countries. In Spain, there is almost an order of magnitude difference for the median $\mathcal{E}$ between the least and the most precise ISPs. For the French case, this difference is even broader, with SFR as the best option with a remarkably high median precision of 4.5Km, and Orange as the worst case for GeoIP location purposes with $\mathcal{E}$=173.2Km in median. Considering that both operators are the most popular in France, according to their popularity in the \predicio dataset, we ascribe this difference to a worse performance of the position matching algorithm used by \netacuity. Finally, the ISP choice in GB has the lowest impact among the analyzed countries, with a close gap between the median $\mathcal{E}$ of the analyzed operators. In a nutshell, despite few cases, for all the analyzed operators, the 25th percentile of $\mathcal{E}$ is above 10Km. This confirms that making a fine-grain selection of GeoIP locations per operator cannot be used for precise location-targeted advertising or other similar services.

\subsubsection{Accuracy}
\label{subsub:acc_isps}

We measure $\mathcal{A}$ for the different carriers in Fig~\ref{fig:accuracy_carrier}. As expected, the carriers that yield a lower median $\mathcal{E}$ generally translate into a higher $\mathcal{A}$. However, the quite large differences in the median precision observed in France do not translate into very large differences in terms of accuracy, while the less dispersed situation in GB yields to a quite diverse performance for some carriers, especially for the Level 2 divisions. Instead, the differences in Spain in terms of $\mathcal{E}$ have a more direct relationship to $\mathcal{A}$, as \netacuity reaches the lowest values for the Yoigo operator. This hints at the complexity of the task GeoIP databases perform: a complex mix of active and passive measurements that \emph{blackbox} the core networks and the interconnections of ISPs.

\begin{figure*}[t!]
    \centering
    \includegraphics[width=\linewidth]{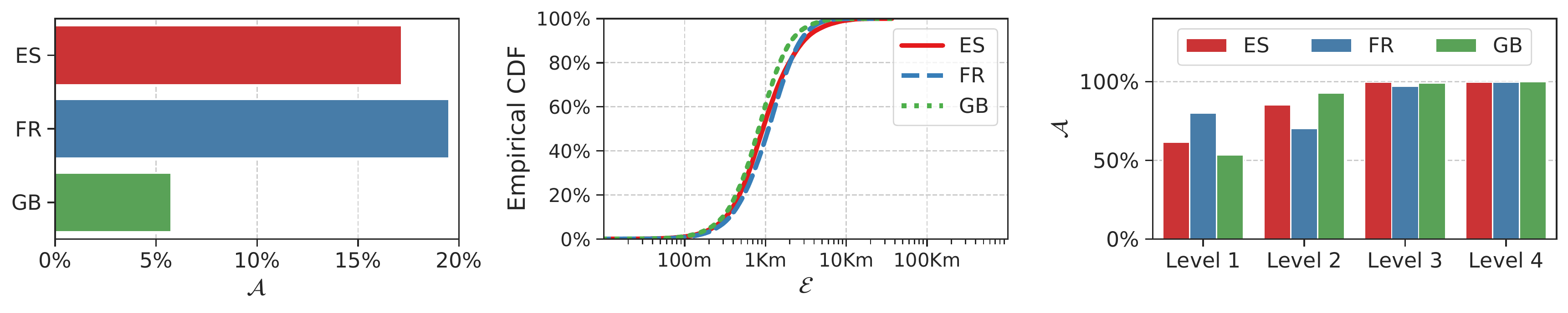}
    \vspace{-6mm}
    \caption{$\mathcal{A}$ for \netacuity to the best possible  \emph{anchor point} (left), $\mathcal{E}$ (center) and $\mathcal{A}$ (right) achieved by the best possible scenario.} 
    \label{fig:acc_voronoi}
\end{figure*}

\subsection{Providers' reported performance}
\label{subsub:maxmind_validation}

Most GeoIP Databases provide high-level reporting about the offered precision and/or accuracy except for Maxmind, that offers a detailed reporting~\cite{maxmind_report}. Despite Maxmind’s report does not cover as many dimensions as we cover in our research, it offers precision data at several thresholds (10, 25, 50, 100, and 250km) and accuracy values at two administrative levels (zip code and city) for around a hundred countries. An important difference with Maxmind is that we provide a detailed description of our methodology to study the precision and accuracy of \new{GeoIP}, whereas Maxmind does not disclose their methodology. 

We have compared Maxmind’s and our outcome for the three analyzed countries in this paper. The results are, in general, well-aligned. This means our study is the first academic validation of the correctness of the precision and accuracy results reported by Maxmind. 

\section{Dissecting the GeoIP internals}
\label{sec:voronoi}

In this section, we go one step further and try to analyze the different components that may contribute to the lack of performance analyzed in \S \ref{sec:accuracy}. \new{Both \netacuity and \maxmind do not disclose the algorithm and techniques they use to provide the mapping between IP and location, although we know from generic statements published on the vendor websites and from the literature~\cite{} that they are likely using a mixture of active and passive measurement, possibly combined with machine learning technologies and datasets close to \predicio.

While proposing improved solutions for \new{GeoIP} is out of the scope of this paper, in this section we propose a methodology to discover the upper bound of the GeoIP performance, a metric that we will leverage for the online advertising case study discussed in \S \ref{sec:advertising}.
}

\subsection{Methodology}
\label{sub:voronoi}

Mapping an IP address to its $pos_{IP}$ as performed by the two databases, is a function internally composed by two sub-tasks; $i$) create a grid of possible \emph{anchor points}, based on passive and active measurements, and $ii$) assign each IP prefix to such grid. While these two operations are likely to be conducted together, thus creating a ```moving'' map of the anchor points \mg{which may have considerable drifts, as reported in~\cite{gouelip}}, in this section, we analyze the precision of the technology by splitting these two tasks into two independent phases, assuming that the mapping is performed on top of a set of already defined anchor points. While this operation is not feasible for \netacuity and \maxmind, it serves as a \emph{best case} for GeoIP.

To this end, we take the set of $pos_{IP}$ shown in Fig.~\ref{fig:map} in each of the countries and compute its Voronoi tessellation~\cite{voronoi}, actually defining the areas in which the errors are minimized. We measure the performance of the GeoIP databases using the same \mg{metrics} discussed in \S \ref{sub:accmetho}: $\mathcal{E}$ and $\mathcal{A}$. We next analyze this scenario under different configurations.

\subsection{Internal accuracy}
\label{sub:intaccuracy}

The first question that we want to answer is how frequently the mapped $pos_{IP}$ is actually the best possible one.  \new{We answer this question by computing the fraction of users that are mapped (with both $pos_{GT}$ and $pos_{IP}$ within the same Voronoi cell. If this happens, then it means that the selected $pos_{IP}$ is actually the \emph{best} possible one among the set of available anchor points. If not, it means that there was an anchor point closer to $pos_{GT}$ than $pos_{IP}$ that was not selected. The results of this analysis are shown in the left part of Fig.~\ref{fig:acc_voronoi}}

In this situation, \netacuity cannot go beyond an overall accuracy above 20\% in the best case (Spain), i.e., more than 80\% of the IP addresses are not mapped to the \emph{best} location. This is even more dramatic for the GB case, where just 6\% of the addresses are mapped to the best anchor points. 

\new{This corroborates the complexity of the tasks that GeoIP providers face: while they can quite effectively map more densely populated areas with more anchor points, the users' geographical spread (especially for the ones using the mobile network, as we analyze in \S \ref{subsub:prec_access}) makes very difficult to condensate IP ranges into the best possible anchor point.}

\subsection{Anchor points optimal granularity}
\label{sub:optimal}

In this experiment, we go one step forward and analyze a \emph{what if} scenario in which we assume that the mapping to the $pos_{IP}$ is dynamically performed at each query over a fixed grid of predefined anchor points (i.e., the ones already present in \netacuity) \mg{selecting the \emph{best} possible anchor point, i.e., the one with the least euclidean distance to $pos_{GT}$}. This approach allows us to remove any possible error due to the end users movement, so that we can evaluate the quality of the measurements performed to map a given prefix onto a $pos_{IP}$. That is, the \netacuity developers could find a prefix whose unique characteristics are uniquely identified through their algorithms by a \texttt{lat,long} pair, showing the potential of their system.

We evaluate this fact by repeating the analysis we performed in \S \ref{sub:variance}, assessing the improvement in terms of $\mathcal{E}$ and $\mathcal{A}$. Results are shown in Fig.~\ref{fig:acc_voronoi}. Improvements are dramatic: 99.9\% of all our samples fall below the 10Km $\mathcal{E}$ mark, a remarkable precision that could be a game-changer for many applications, including online advertising. Indeed, $\mathcal{A}$ also grows towards the highest accuracy levels, with the precision for Level 1 (the most challenging one) that grows from around 10\% (see Fig.~\ref{fig:accuracy}) to more than 60\%, being practically \emph{error free} at Levels 3 and 4.

These results show that $i$) \netacuity technology can accurately measure where users actually are at a coarser granularity, as the \predicio population always has an \emph{anchor point} within 10 Km in the vast majority of cases, but $ii$) end users micro-mobility largely spoils the achieved granularity. We claim that if GeoIP providers were able to account for this micro-mobility in their mapping, the impact on the final applications such as online advertising would be huge, as we discuss in the following section. 

\section{The impact of GeoIP on online advertising}
\label{sec:advertising}

This section aims to estimate what is the impact of using GeoIP locations on online advertising campaigns. While the extensiveness of the \predicio dataset used in \S \ref{sec:accuracy} allowed us to understand the performance of the GeoIP overall, that dataset  may include all kinds of users, not only the ones which are actively targeted by ad providers. To this end, we use our bid stream dataset to generate the ground-truth data to guarantee that all the location samples are actually linked to users targeted by online advertising campaigns. 

To measure the performance of a campaign, we rely on the accuracy ($\mathcal{A}$) measure described in \S \ref{sec:accuracy}. However, to understand the best buying strategy from an economic point of view, in addition to the accuracy, we also have to consider the monetary cost $\mathcal{C}$ associated with different types of bid-requests, i.e., including GeoIP or GPS information.
In order to isolate the monetary impact that the type of location data has on advertising campaigns, we need to factor out other elements affecting the economic performance of a campaign. To this end, we make the following assumptions: $i$) the bid stream has been filtered so that the available bid-requests already meet the goals of the campaign in terms of the targeted audience; $ii$) there is sufficient ad inventory of each type of location data (GeoIP vs. GPS) to meet the defined objective of the campaign in terms of the number of ad impressions delivered, so that the advertiser/DSP can freely choose to buy any combination of GeoIP and GPS bid-requests to meet such objective.

\subsection{Methodology}

\subsubsection{Best bidding strategy}
\label{subsub:strategy}

\new{In this section, we model the best strategy that could be followed by an advertiser to issue a specific targeting ad campaign, based on the characteristics of the location technology and their associated cost. The goal of the advertiser is to maximize the \emph{value for money} for every ad campaign. Let us introduce this in a toy example, where the GPS accuracy is 100\% by definition and the GeoIP accuracy of the bid requests in this campaign is 20\% (i.e., the location of the targeted user matches the location defined by the ad campaign once every five times). In this case, if the average cost of the GPS bid request is twice the cost of the GeoIP bid request, it would be more economically effective to buy GPS bid requests. However, if the cost of GPS bid request was 6 times the cost of GeoIP bid requests, it would be more economically effective to buy the latter.

To model this behaviour, we} introduce the normalized cost related to each technology  ($\mathcal{C}^*_{IP}$ and $\mathcal{C}^*_{GPS}$), computed as: 
\begin{align}
 \mathcal{C}^*_{IP}  &= \frac{\mathcal{C}_{IP}}{min\left(\mathcal{C}_{IP},\mathcal{C}_{GPS}\right)} & \mathcal{C}^*_{GPS} &= \frac{\mathcal{C}_{GPS}}{min\left(\mathcal{C}_{IP},\mathcal{C}_{GPS}\right)}
 \label{eqn:cost}
\end{align}
Then, by using the accuracy obtained with the GeoIP and GPS technologies ($\mathcal{A}_{IP}$ and $\mathcal{A}_{GPS}$) we can calculate the \emph{Effective Cost} ($\phi$), that is defined as the normalized cost of correctly delivering an ad to a user located in the targeted area, and is calculated as follows:

\begin{align}
 \phi_{IP}  &= \frac{\mathcal{C}^*_{IP}}{\mathcal{A}_{IP}} & \phi_{GPS} &= \frac{\mathcal{C}^*_{GPS}}{\mathcal{A}_{GPS}}
 \label{eqn:phi}
\end{align}

Hence, the best expenditure strategy is defined by the min$\left(\phi_{IP},\,\phi_{GPS}\right)$. For a given location-targeted campaign, by estimating the accuracy and cost for the two technologies, a DSP can steer its strategy according to this rule.  Later in this section,  we empirically evaluate  $\mathcal{A}$ as well as  $\phi$ for different real-world ad campaign scenarios.

Note that our methodology is not considering the potential economic side benefits/harms of showing ads to users outside the targeted area. For instance, a potential benefit might be expanding the knowledge of a new brand to neighboring areas of the specific location target. Instead, potential harm might be bothering users with ads uninteresting to them, which in addition introduces a waste of resources (e.g., bandwidth \cite{ad_waste} and battery).

\subsubsection{Bid stream ground-truth dataset}
\label{subsub:gt-dataset}

In order to precisely measure $\mathcal{A}_{IP}$, we have to select a set of bid-requests for which we know the end users ground-truth location.
We do this by keeping exclusively the bid-requests that include a GPS location (See \S \ref{subsec:loc_data_back}), hence creating a reliable association between the users' IP addresses and their position $pos_{GT}$ (we assume that $\mathcal{A}_{GPS}=100\%$). Then, we retrieve the location information from the Geolocation databases using the IP address, obtaining $pos_{IP-A}$ and $pos_{IP-B}$.  Our ground-truth dataset includes the following information: \texttt{$<$timestamp; IP address;$pos_{GT}$;$pos_{IP-A}$;$pos_{IP-B}$$>$}. 

Moreover, in order to retrieve the value of $\mathcal{C}_{GPS}$ and $\mathcal{C}_{IP}$ we rely on the bid floor information available in our bid stream dataset. Note that for estimating $\phi_{GPS}$ and  $\phi_{IP}$ the relevant information is not the absolute price value for GPS and GeoIP ad impressions but the relative relation between them ($\mathcal{C}^*_{GPS}$ and $\mathcal{C}^*_{IP}$). Hence, our assumption here is that the ratio of GeoIP and GPS price value is well captured by the ratio of their corresponding bid floor prices.

\subsubsection{Simulation set-up}
\label{subsub:simulation}

Our goal is to create a simulation set-up that mimics real location-targeted ad campaigns. For this purpose, we follow the guidance from industry players, such as TAPTAP Digital, to set up realistic values for our simulation parameters as described next: 

\noindent \textbf{Campaign duration}: We set up a campaign duration between 1 and 2 weeks, which is a very common time frame used by advertisers for their ad campaigns.
    
\noindent \textbf{Win rate}: This parameter defines the fraction of won bid-requests out of all the bids run by a DSP in an ad campaign. We configure a win rate range between 20 and 40\% in our reference ad campaigns.
    
\noindent \textbf{Ad impression cost}:  We use the \emph{bid floor} as a proxy metric to estimate the cost of ad impressions. We have computed the $C^*_{IP}$ and $C^*_{GPS}$ defined above as the median value of bid floor prices for GeoIP and GPS bid-requests collected for Spain, France, and Great Britain across 16 days. $C^*_{IP}$ is 1 for the three countries, whereas $C^*_{GPS}$ is 1.01, 2.34, and 2.08 for Spain, France, and Great Britain, respectively.

\noindent \textbf{Geographical target}: We consider campaigns targeting all administrative levels introduced in Table \ref{tab:admin_levels} but the country level. As discussed in \S \ref{sec:accuracy}, GeoIP services have perfect accuracy in providing the location at country level. Then, it is expected that country level campaigns have an $\mathcal{A}~\approx$~100\%. Note that the 4 levels used in our simulations (state, province, city, and zip code) are frequently used as targeted-locations in online advertising campaigns.

\noindent \textbf{Urbanization level}: A major portion of location-targeted advertising campaigns focus on urban areas. Then, our simulations will focus on this type of areas. Note that the urbanization level is only meaningful for administrative Levels 1 (zip code) and 2 (city) since we cannot select a province or a state which is entirely urban or rural.

For each of the considered countries (Spain, France, and Great Britain), we configure 4 campaign models based on the geographical target and urbanization level: a) Level 4, b) Level 3, c) Level 2-Urban, and d) Level 1-Urban. Overall, we have a total of 12 simulation scenarios. For each simulation scenario, we randomly select 5 different targets that meet its criteria, with the exception of Great Britain, which does not account for Level 4  as it is in general yielding very high accuracy (see Fig.\ref{fig:accuracy}), hence generating 3 total targets.

Overall, we have 58 different target-locations in our stimulation set.  Finally, for each of the 58 campaigns, we run 3 repetitions where we set up a value of campaign duration and win rate randomly selected from the range defined above for these parameters. 

\subsubsection{Campaign execution}
\label{subsub:simprocess}

We execute the simulated campaigns on the bid stream coming from our ground-truth dataset.
We filter only the bid-requests, including a $pos_{IP-A}$ or $pos_{IP-B}$ location matching the geographical target of the ad campaign in the selected time period. We just consider a random fraction of the bids from the obtained subset according to the win rate defined for the campaign. The final set of bid-requests resulting from this process represents the actual set of delivered ad impressions by the ad campaign.  

\subsubsection{Evaluation metrics}
\label{sub:evaluation}

First, we compute the Accuracy ($\mathcal{A}_{IP}$) metric to assess the impact of GeoIP location data in online advertising. We measure the accuracy on the delivered ad impressions as the fraction of them whose associated $pos_{GT}$ falls within the specific geographical target of the ad campaign. For each of the 58 simulated campaigns, we compute the average $\mathcal{A}_{IP}$ across the three performed repetitions. Note that as indicated above, $\mathcal{A_{GPS}}$= 100\%.

Second, we compute $\phi_{IP}$ vs. $\phi_{GPS}$ using the expressions defined in Eq. \ref{eqn:phi} to identify the technology (GeoIP or GPS) yielding the most economically efficient campaign.

Third, using the values of $\phi_{IP}$ and $\phi_{GPS}$, we define the \emph{Gain} ($\mathcal{G}_{IP}$) of an ad campaign, as follows:

\begin{equation}
\centering
\mathcal{G}_{IP}= log\left(\frac{\phi_{GPS}}{\phi_{GeoIP}}\right)
\label{eqn:G}
\end{equation}
$\mathcal{G}_{IP}$ quantitatively compares the value increase (decrease) in accuracy for the GeoIP with the increase (decrease) in their cost. Thus, positive (negative) values of $\mathcal{G}_{IP}$ provide a quantitative reference of the expected order of magnitude improvement (harm) of setting a strategy to buy  GeoIP instead of GPS bid-requests.

Finally, note that we compute $\mathcal{A_{IP}}$, $\mathcal{A_{GPS}}$, $\phi_{IP}$, $\phi_{GPS}$ and $\mathcal{G}_{IP}$  for both: $i$) the \textit{Actual} mapping of the IP addresses location to the anchor points implemented in \netacuity, and $ii$) the \textit{Optimal} assignment of IP addresses to the closest anchor point, as discussed in \S \ref{sec:voronoi}.

\subsection{Results}
\label{sub:campaign_results}

We note that the results presented in this section correspond to \netacuity. \mg{For the sake of simplicity, we do not report the results associated with \maxmind which lead to the same conclusions.}

\subsubsection{Accuracy}
\label{sub:simaccuracy}

\begin{figure}[t!]
    \centering
    \includegraphics[width=\linewidth]{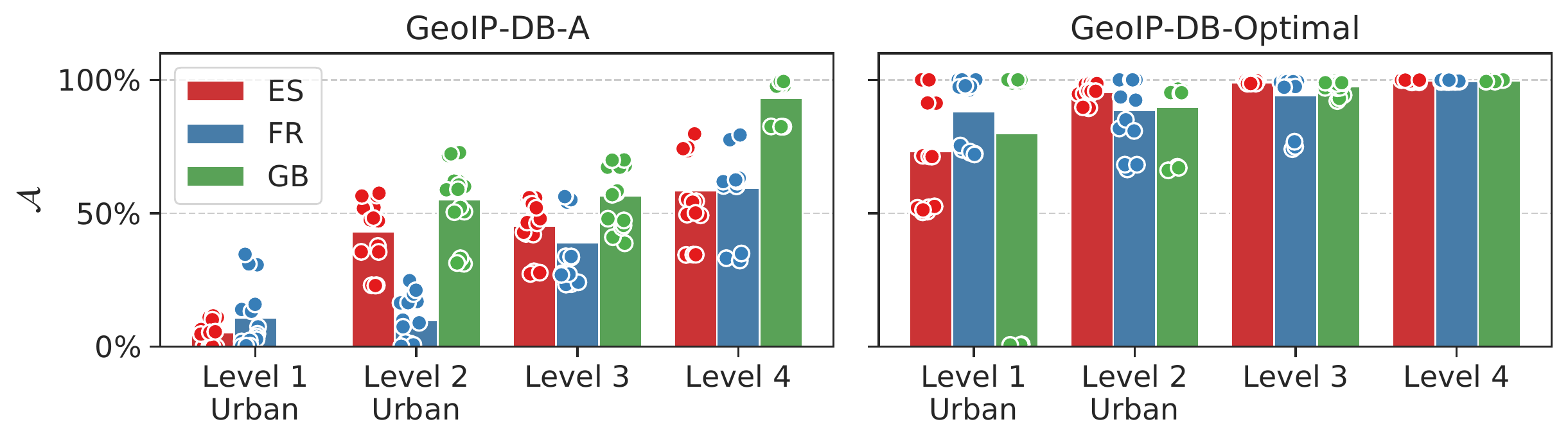}
    \vspace{-5mm}
    \caption{Accuracy by administrative regions}
    \label{fig:accuracy_simulaciones}
\end{figure}

Fig. \ref{fig:accuracy_simulaciones} shows the accuracy from the ad campaigns simulations for the four geographical targets introduced in \S \ref{subsub:simulation} in Spain, France and Great Britain when we consider the \textit{Actual} (left side) or the \textit{Optimal} (right side) mapping of IP addresses location to anchor points.

The results of the \textit{Actual} allocation strategy follow the expected pattern for $\mathcal{A}$: the larger is the geographical target, the higher is the accuracy. Using Spain to illustrate this observation: $\mathcal{A}$ grows from 5.25\% for campaigns targeting zip codes in urban areas to 58.45\% when the campaign resolution is at the state level. 

In addition, it is interesting to notice that the accuracy varies considerably across countries in all the geographical targets, except for Level 3. 
Also, the accuracy reported with the \predicio dataset (See Fig. \ref{fig:accuracy} in \S \ref{sec:accuracy}) shows more evenly spread behavior across countries. This suggests that the users targeted by online advertising can be a rather skewed subset of the overall population that can be reached by high-precision location providers discussed in \S \ref{sec:accuracy}.

When analyzing the \textit{Optimal} assignment of IP addresses to the closest anchor point, we find that it largely outperforms the \textit{Actual} allocation irrespective of the geographical target we consider, as expected. The worst case in the \textit{Optimal} allocation ($\mathcal{A}=73.18\%$) corresponding to the zip code level in urban areas in Spain is only 20 percentage points smaller than the best case in \textit{Actual} allocation algorithm ($\mathcal{A}=93.16\%$), which comes from the sate level in GB. 

In conclusion, the average $\mathcal{A}$ for the \textit{Optimal} allocation strategy yields advantages for all geographical resolutions. If the GeoIP services were capable to approximate this \textit{Optimal} performance, advertisers using location-targeted campaigns would experience a significant improvement in their campaigns' KPIs without requiring any further investment.

\begin{figure}[t!]
    \centering
    \includegraphics[width=\linewidth]{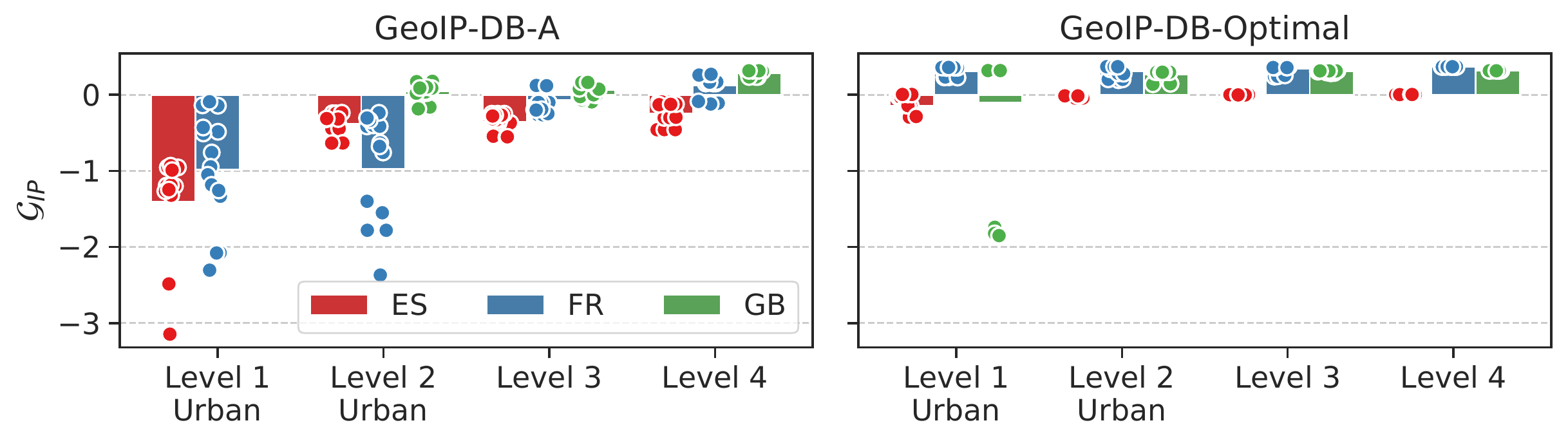}
    \vspace{-5mm}
    \caption{Relative economic gain or loss ($\mathcal{G}_{IP}$)}
    \label{fig:gain}
\end{figure}

\subsubsection{Optimal budget strategy}
\label{sub:budget}

Tab. \ref{tab:ec} shows the best buying strategy (i.e., buying GeoIP vs. GPS bid-requests) to be applied in each of the campaigns run for the four geographical targets introduced in \S \ref{subsub:simulation} for the three countries as a result of comparing $\phi_{IP}$ and $\phi_{GPS}$ in the simulated campaigns. Results are grouped by target location. For each target location, the table shows the number of experiments where the recommended strategy is buying GPS only or if GeoIP is an economically more profitable alternative.

First, if we consider the \emph{Actual} allocation of IP prefixes to anchor points, GeoIP would be the recommended technology in  30\% and 55\% of the experiments in France and Great Britain, respectively, whereas GPS is the recommended technology in all ad campaigns from Spain. This result is a consequence of a higher cost of GPS bid-requests in France and Great Britain, which recommends buying GeoIP over GPS in certain campaigns, even when the accuracy of GeoIP is significantly worse. In the case of Spain, the cost of GPS and GeoIP bid-requests is very similar so, the superior precision of GPS leads to recommend buying this type of bid-requests.

Second, the number of campaigns where GeoIP becomes the best buying strategy when we use the \emph{Optimal} allocation instead of the \emph{Actual} one grows from: 0 to 27 in Spain, from 18 to 60 in France, and from 30 to 51 in Great Britain. This confirms that, if GeoIP providers could further improve their technology to achieve a better approximation of users’ positions, as discussed in \S \ref{sec:voronoi}, this technology has a large potential for further improvements for this application.

To conclude our analysis, we study the budgetary improvement (harm) imposed by a strategy focused on buying only GeoIP bid-requests against the opposite one that only focuses on the ones that carry GPS information. To this end, the main bars in Fig. \ref{fig:gain} show the average $\mathcal{G}_{IP}$ for each considered targeted-location in our ad campaigns, while the points are the $\mathcal{G}_{IP}$ for each specific realization of the ad campaigns\footnote{As $\mathcal{A} \approx 0$ for all experiments targeting Level 1-Urban GB in \netacuity, the figure does not show this configuration.}. Again, we present results considering the \emph{Actual} performance of the GeoIP and the \emph{Optimal} allocation of prefixes to anchor points. Overall, this figure provides a more detailed picture of the findings derived from Tab. \ref{tab:ec} by quantifying the advantage (disadvantage) of having an ad buying strategy that is entirely composed by GeoIP.

First of all, the current price differences that favor the cheaper GeoIP against the more expensive GPS bid-requests limit the maximum $\mathcal{G}_{IP}$ that GeoIP can offer compared to GPS to less than an order of magnitude, even in those cases where GeoIP offers a perfect accuracy. Instead, in the configurations that yield a very poor accuracy using GeoIP (e.g., Level 1-Urban in France for \netacuity), there is a severe economic impact as measured by $\mathcal{G}_{IP}$: the wrong strategy of buying only GeoIP bid-requests can worsen the monetary efficiency by a factor of 10.



The technologies become comparable just when the accuracy grows to very high values, hence the increase in accuracy yielded by the GPS technology is not compensated by the higher cost of GPS bid-requests. Only for the Spanish case, where GPS ads are comparatively cheaper than in other countries in our dataset, GeoIP is still lagging behind from a monetary point of view.

Finally, Fig. \ref{fig:gain} also shows $\mathcal{G}_{IP}$ in the case of the \emph{Optimal} assignment. Comparatively, with such an improvement, GeoIP would be on par with GPS in 3 cases, and even yielding a constant monetary gain in 9 of the 12 scenarios.

\begin{table}[t!]
\resizebox{\columnwidth}{!}{%
\begin{tabular}{@{}llrr@{}}
\toprule
\multicolumn{1}{c}{\textbf{Country}} & \multicolumn{1}{c}{\textbf{Target Location}} & \multicolumn{1}{c}{\textbf{\begin{tabular}[c]{@{}c@{}}Best Tech. \\ Actual\end{tabular}}} & \multicolumn{1}{c}{\textbf{\begin{tabular}[c]{@{}c@{}}Best Tech.\\ Optimal\end{tabular}}} \\ \midrule
ES                                   & Level 1 (Urban)                              & (GPS, 15)                                                                                 & (GPS, 12), (GeoIP, 3)                                                                   \\
ES                                   & Level 2 (Urban)                              & (GPS, 15)                                                                                 & (GPS, 15)                                                                                 \\
ES                                   & Level 3                                      & (GPS, 15)                                                                                 & (GPS, 6), (GeoIP, 9)                                                                    \\
ES                                   & Level 4                                      & (GPS, 15)                                                                                 & (GeoIP, 15)                                                                                  \\ \midrule
FR                                   & Level 1 (Urban)                              & (GPS, 15)                                                                               & (GeoIP, 15)                                                                                 \\
FR                                   & Level 2 (Urban)                              & (GPS, 15)                                                                               & (GeoIP, 15)                                                                       \\
FR                                   & Level 3                                      & (GPS, 9), (GeoIP, 6)                                                               & (GeoIP, 15)                                                                               \\
FR                                   & Level 4                                      & (GPS, 3), (GeoIP, 12)                                                                            & (GeoIP, 15)                                                                               \\ \midrule
GB                                   & Level 1 (Urban)                              & (GPS, 15)                                                                                & (GPS, 3), (GeoIP, 12)                                                                \\
GB                                   & Level 2 (Urban)                              & (GPS, 3), (GeoIP, 12)                                                              & (GeoIP, 15)                                                                               \\
GB                                   & Level 3                                      & (GPS, 6), (GeoIP, 9)                                                                    & (GeoIP, 15)                                                                               \\
GB                                   & Level 4                                      & (GeoIP, 9)                                                                               & (GeoIP, 9)                                                                               \\ \bottomrule
\end{tabular}}
\caption{Best technology (GPS vs. GeoIP) to set the buying strategy based on the analysis of $\phi$.}
\label{tab:ec}
\end{table}

\section{Related Work}
\label{sec:relwork}

The utilization of IP addresses as a proxy for geolocating devices has attracted the interest of the research community for more than a decade now, showing the importance of the topic due to the widespread use of GeoIP solutions in online advertising, fraud detection or anti-piracy solutions. 

A first body of work, which does not directly study the performance of GeoIP,  analyzed the geographical allocation of IP addresses and IP prefixes from both spatial and temporal angles. Gueye et al. \cite{10.1007/978-3-540-71617-4_26} reported the difficulty of accurately geolocate an IP address due to the geographic span of IP addresses blocks. Almost 15 years later, and with a huge proliferation of the utilization of GeoIP location services, these findings seem to remain valid. Padmanabhan et al. \cite{10.1145/3386367.3431314} studies the duration of IPv4 and IPv6 assignments to a device through large-scale measurements. The paper shows that 75\% of mobile devices get IP addresses assigned for a duration lower than a day, whereas devices with a fix connections keep the same IP address for dozens of days typically. An earlier work by Balakrishnan et al. \cite{10.1145/1644893.1644928} performs a similar measurement study on the stability of IP address assignment in early 3G mobile networks in US. The paper reports that the IP addresses used by a mobile device could change in few minutes and thus they do not embed geographical information with enough granularity to implement GeoIP solutions based on them. These results support our finding that the error of GeoIP-based locations of IP addresses (or prefixes) using cellular access connections is significantly larger that those using fixed connections.

The closest literature to our study is formed by studies that analyze the accuracy of GeoIP databases. In one of the earliest studies on the topic, Poese et al. \cite{poese2011ip} use data from an ISP to analyze the performance and accuracy of 5 different GeoIP databases. In particular, they find that none of these databases make a good mapping of the actual IP prefixes used by the ISP. Furthermore, they also map the location of each IP prefix to the location of the Point-of-Presence (PoP) where the associated backbone router is located. Unfortunately, this location ground-truth might be significantly less accurate than GPS coordinates from a mobile device as we use in this paper. In an almost parallel study in time, Shavitt at al. \cite{6081357} compare the performance of 6 GeoIP databases. They use two types of ground-truth datasets: the geographical location of PoPs and a ground-truth database, including the location of 25k IP address up to the level of city. The paper uses the precision as the studied performance metric. The authors also analyze the correlation between the error of different GeoIP databases.

There are very few papers in the literature using ground-truth data based on GPS location information. Triukose et al. \cite{triukose2012geolocating} leverage the GPS location provided by a mobile app, and assess the error of GeoIP location services using the IP address of the device. Complementary to our study, this paper shows evidence that NATed IP addresses offer a worse location accuracy than public IP addresses. However, this study present an important limitation since the dataset only include information about devices connected through cellular (3G/GPRS) technology. In a similar study, Komosny et al. \cite{komosny} use 700 mobile devices from which they recover the GPS location to construct a ground-truth dataset to evaluate the performance of 8 different GeoIP databases. While, these studies rely on GPS ground-truth data, their dataset is formed by tens of thousands of location samples compared to more than 2B samples in our dataset. 

Finally, there are some previous works complementary to ours, which analyze the performance of GeoIP databases in geolocating network infrastructure elements. Instead, we are interested in analyzing the performance in the geolocation of end users. Gharaibeh et al. \cite{10.1145/3131365.3131380}, use a ground-truth dataset including the city level location of 16.5K router interface IP addresses, whereas Iordanou et al. \cite{nikos} focus their analysis on the location of servers. Both works conclude that GeoIP databases are highly inefficient in geolocating network infrastructure elements.

Our study presents three major contributions in comparison with the previous literature:  1) we present the first benchmark analysis about the upper-bound performance that GeoIP could offer (see \S \ref{sec:voronoi}); 2) To the best of the authors’ knowledge, all existing works analyze the GeoIP databases performance in an isolated manner and just briefly mention which businesses might be affected by the reported inaccuracy of GeoIP. Instead, we present, for the first time, a detailed quantitative analysis of the potential impact of the extensive use of GeoIP in online advertising, which arguably represents the most important business where GeoIP is applied; 3) We present the most thorough study of GeoIP performance in terms of scale and resolution. In particular, our study leverages over 2B ground-truth samples with GPS precision. This allows us to present the most comprehensive study of the GeoIP performance, studying it up to a zip code resolution and covering the impact of several relevant factors such as the level of urbanization, the access technology or the specific ISP.

As a final remark, to the best of the authors’ knowledge, there is only one company, Location Sciences \cite{locationsciences}, offering location data auditing products in the online advertising ecosystem. Unfortunately, as all other auditing solutions in online advertising \cite{dv,human,ias} their products are proprietary and it is unknown how they operate or which is their actual performance.

\section{Conclusion}
\label{sec:conclusion}

To the best of the authors’ knowledge, our study is: 1) the one that provides a deepest understanding of the performance of GeoIP databases; 2) the first one providing an upper bound of the performance these systems may offer and 3) The first one analyzing its impact on the online advertising business. These three elements constitute (in our humble opinion) an important contribution to researchers and practitioners and make our paper novel compared to any other previous study in the context of GeoIP databases.

\new{In this paper, we present an analysis of two GeoIP databases, that are arguably among the most widespread technologies used to locate devices around the entire world, especially in the context of online advertising. To the best of our knowledge, our study is: $i$) the one that provides the deepest understanding of the performance of GeoIP databases; $ii$) the first one providing an upper bound of the performance these systems may offer, and $iii$) the first one analyzing its impact on the online advertising business.}

Armed with a dataset of 2B samples that includes a ground-truth location associated with an IP address, we study the performance of \new{GeoIP} databases through several unexplored dimensions so far: urban vs. rural areas, access technologies, or ISP providers. Our work revisits the quantitative findings of previous studies regarding the performance issues of this technology and extends them to understand their causes better.

Thanks to the extensiveness of our data, we can further dig into the performance of \new{GeoIP}  databases, showing possible causes behind the lack of accuracy and discussing how, under \emph{ideal} conditions, the overall precision could be improved by two orders of magnitudes.

Finally, we prove that from a budgetary perspective, GeoIP may be, in some cases, a better technology for geographically targeted ad campaigns compared to more precise geolocation technologies (i.e., GPS) due to the expected higher cost of the latter. The most efficient technology in economic terms is the one that better balances accuracy and cost. This is initially a counter-intuitive result since most of the literature in the area mostly focuses on reporting the poor location capacity of \new{GeoIP} databases. 

\section*{Acknowledgements}

This research received funding from the European Union’s Horizon 2020 innovation action programme under the PIMCITY project (Grant 871370) and the TESTABLE project (Grant 101019206); the Agencia Estatal de Investigación (AEI) under the ACHILLES project (Grant PID2019-104207RB-I00/AEI/10.13039/501100011033); the Spanish Ministry of Economic Affairs and Digital Transformation and the European Union-NextGenerationEU through the UNICO 5G I+D 6G-RIEMANN-FR; the agreement between the Community of Madrid and the Universidad Carlos III de Madrid for the funding of research projects on SARS-CoV-2 and COVID-19 disease, project name "Multi-source and multi-method prediction to support COVID-19 policy decision making", which was supported with REACT-EU funds from the European regional development fund “a way of making Europe; and the TAPTAP-UC3M Chair in advanced AI and Data Science applied to advertising and marketing.

\bibliographystyle{IEEEtran}
\bibliography{IEEEabrv,tmc21_geoipaccuracy}

\begin{IEEEbiography}[{\includegraphics[width=1in,height=1.25in,clip,keepaspectratio]{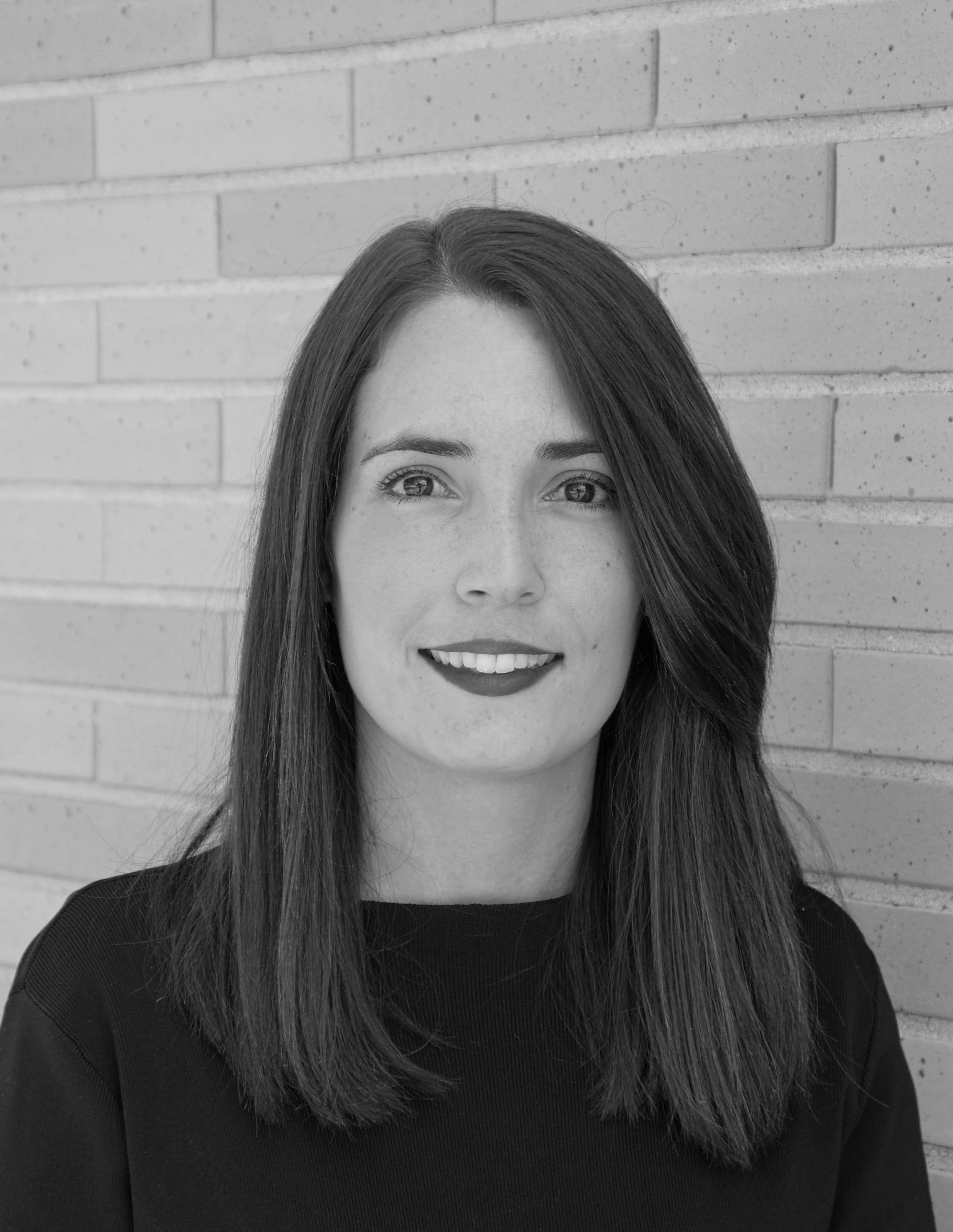}}]
{Patricia Callejo} is a post-doc researcher at UC3M-Santander Big Data Institute. She obtained her M.Sc (2016) and Ph.D (2020) at University III of Madrid in the field of Telematics Engineering. She was granted by RIPE Academic Cooperation Initiative (RACI) on RIPE 76 that took place in Marseille, France in 2018. The same year, she did an internship in the International Computer Science Institute (ICSI) at UC Berkeley (USA), as part of her PhD. She is author of conference papers such as ACM HotNets, ACM CoNEXT, and WWW. She has participated in EU H2020 projects. Her areas of interest include Internet measurements, online advertising, and web transparency.
\end{IEEEbiography}

\begin{IEEEbiography}[{\includegraphics[width=1in,height=1.25in,clip,keepaspectratio]{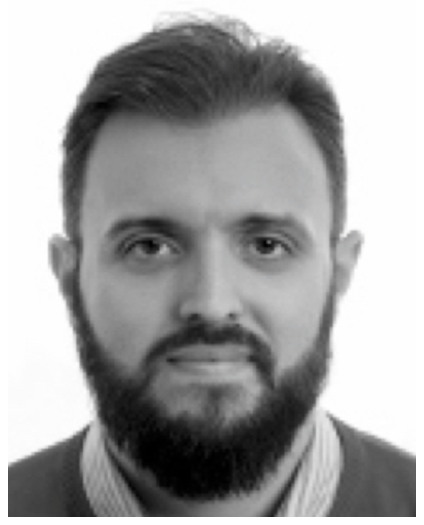}}]
{Marco Gramaglia} 
received the M.Sc. and Ph.D.
degrees in telematics engineering from the University Carlos III of Madrid (UC3M), in 2009 and
2012, respectively. He held post-doctoral research
positions at ISMB, Italy, CNR-IEIIT, Italy, and
IMDEA Networks, Spain. He is currently a PostDoctoral Researcher at UC3M. He was involved in
EU projects and has authored more than 50 papers
appeared in international conference and journals.\end{IEEEbiography}

\begin{IEEEbiography}[{\includegraphics[width=1in,height=1.25in,clip,keepaspectratio]{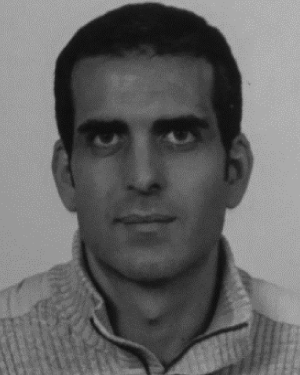}}]
{Rub\'en Cuevas} is an Associate Professor in the Telematic Engineering department at Universidad Carlos III de Madrid, Spain and the Deputy Director of the UC3M-Santander Big Data Institute.  He has coauthored  over 70 papers in prestigious international journal and conferences such as ACM CoNEXT, WWW, Usenix Security, ACM HotNets, the IEEE Infocom, ACM CHI, IEEE/ACM TON, the IEEE TPDS, CACM, PNAS, Nature Scientific Reports, PlosONE, or Communications of the ACM. He has been the PI of 10 research projects and participated in more than 25 projects. His research work has been featured in major internationalca media such as The Financial Times, BBC, The Guardian, New Scientist, Wired, Corriere della Sera, Le Figaro, El Pais, etc. His main research interests include online advertising, web transparency and Internet measurements.
\end{IEEEbiography}

\begin{IEEEbiography}[{\includegraphics[width=1in,height=1.25in,clip,keepaspectratio]{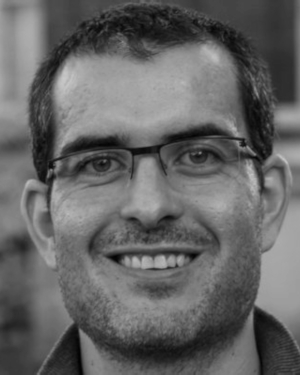}}]
{\'Angel Cuevas} received the M.Sc. (2007), and
the Ph.D.(2011) degrees in Telematics Engineering from
the University Carlos III of Madrid. He is currently an Associate Professor in the Department of Telematic Engineering, University Carlos III of Madrid. He is a co-author of more than 70 papers in prestigious international journals and conferences, such as the IEEE/ACM TRANSACTIONS ON NETWORKING, the ACM Transactions on Sensor Networks, Computer Networks (Elsevier), the IEEE NETWORK, the IEEE Communications Magazine, USENIX Security, WWW, ACM CoNEXT, and ACM CHI. His research interests focuses on Internet measurements, web transparency, privacy, and P2P networks. He was a recipient of the Best Paper Award at ACM MSWiM 2010.
\end{IEEEbiography}

\end{document}